\documentclass{aa}  
\usepackage{graphicx}
\graphicspath{{png/}}
\bibpunct{(}{)}{;}{a}{}{,} 
\usepackage{txfonts}
\usepackage{starfont} 
\usepackage{tabularx}
\usepackage{booktabs}
\usepackage[output-decimal-marker={.},multi-part-units=brackets]{siunitx}[=v2]
\usepackage{Abbr_SIUNITX}
\usepackage{multirow}
\def\ic {IC~\num{485}$\,$}
\def\phaseref {J$0802+2509\,$}
\def\water {\element{H_{2}O}$\,$}

\usepackage{natbib}
\begin{document}

   \title{IC~\num{485}: a new candidate disk-maser galaxy at $\sim$~100 Mpc distance }
\subtitle{Milliarcsecond resolution study of the galaxy nucleus and of the \water megamaser }


   \author{E. Ladu
          \inst{1,2}
          \and
          A. Tarchi\inst{2}
          \and
          P. Castangia\inst{2}
          \and 
          G. Surcis\inst{2}
          \and
          J.A. Braatz\inst{3}
          \and
          F. Panessa\inst{4}
          \and
          D. Pesce\inst{5}
          }

   \institute{Dipartimento di Fisica, Università degli Studi di Cagliari, S.P.Monserrato-Sestu km 0,700, I-09042 Monserrato (CA), Italy\\
              \email{elisabetta.ladu@inaf.it}
         \and
             INAF-Osservatorio Astronomico di Cagliari, via della Scienza 5, 09047, Selargius (CA), Italy
           \and
           National Radio Astronomy Observatory,  520 Edgemont Road, Charlottesville, VA 22903, USA 
           \and
           INAF – Istituto di Astrofisica e Planetologia Spaziali, via Fosso del Cavaliere 100, I-00133 Roma, Italy
           \and
           Harvard-Smithsonian Center for Astrophysics, 60 Garden Street, Cambridge, MA 02138, USA
            }

   \date{Received 24 08 2023; accepted 11 10 2023}

 
  \abstract
   {
   Masers are a unique tool to investigate the emitting gas in the innermost regions of active galactic nuclei and to map accretion disks and tori orbiting around supermassive black holes. IC$\,485$, which is classified as a LINER or Seyfert galaxy, hosts a bright \water maser whose nature is still under debate. Indeed, the maser might be either a nuclear disk maser, a jet/outflow maser, or even the very first example of a so-called  `inclined water maser disk'. 
   }
   {
   We aim to clarify and to investigate the nature of the \water maser in \ic by determining the location and the distribution of the maser emission at milliarcsecond resolution and by associating it with the main nuclear components of the galaxy. In a broader context, this work  might also provide further information for better understanding the physics and the disk/jet geometry in LINER or Seyfert galaxies.
   }
   { 
   We observed the nuclear region of \ic in continuum and spectral-line mode with the Very Long Baseline Array (VLBA) and with the European VLBI Network (EVN). 
  Here, we report multi-epoch (six epochs) and multi-band (three bands: L, C, and K) observations, taken in 2018, with linear scales from $\sim$ 3 to 0.2 pc. 
   }
   {
   We detected two \SI{22}{\GHz} \water maser components separated in velocity by \SI{472}{\km \per \s}, with one centred at the systemic velocity of the nuclear region of IC~\num{485} and the other at a red-shifted velocity.  We measured for the first time the absolute positions of these components with an accuracy better than one milliarcsecond. Under the assumption of a maser associated with an edge-on disk in Keplerian rotation, the estimated enclosed mass is $M_{BH} = 1.2 \times 10^{7} M_{\text{\Sun}}$, consistent with the expected mass for a SMBH in a LINER or Seyfert galaxy.
   Continuum compact sources have also been detected in the nuclear region of the galaxy, although at low level of significance.
   }
   {
   The linear distribution of the detected maser components and a comparison with the high sensitivity single dish spectrum
   strongly suggest that the bulk of the maser emission is associated with an edge-on accretion disk. 
   This makes \ic a new candidate for a disk-maser galaxy at the distance of 122 Mpc.  
   In particular, thanks to the upcoming radio facilities (e.g., the Square Kilometer Array and the next generation Very Large Array) \ic will play an important role, as other sources at similar distances, in our understanding of active galactic nuclei in an unexplored volume of Universe.
   }


\titlerunning{IC$\,485$: a new candidate disk-maser galaxy }
\authorrunning{Ladu E. et al.}

   \maketitle


\section{Introduction}

The most common maser emission line, observed for the first time in \num{1969}~\citep{Cheung1969_1detezMaser}, arises from the \water roto-transitional levels $6_{16}$ and $5_{23}$ and it is emitted at \SI{22}{\GHz}, thus in the radio domain.
Among  the extra-galactic \water maser sources, those with an isotropic luminosity $L_\mathrm{iso} > \num{10}\,L_{\text{\Sun}}$ are traditionally defined megamasers, however, this threshold should be used with caution (see Sect.\,4.2 in~\citealt{Tarchi2011_KiloMaser}). These sources are generally found in active galactic nuclei (AGNs).  The majority of \water masers have been found, so far, associated with radio-quiet AGNs in the local Universe ($z\leq0.05$), classified as Seyfert \num{2} (Sy\num{2}) or Low Ionisation Nuclear Emission-line Regions (LINERs), although with some exceptions \citep[e.g.,][]{Tarchi2012,braatz2018IAUS}. 

The activity of \water maser emission in AGNs have been associated with three main  AGN components: (i) disk; (ii) jet; (iii) nuclear outflows.
Jet and outflow masers arise from the interaction between the jet(s)/outflow(s) and the encroaching molecular clouds or because of an accidental overlap along the line of sight between a warm dense molecular cloud and the radio continuum of the jet/outflow(s). 
Jet maser sources provide important information about the evolution of jets and their hotspots (some examples are NGC~\num{1068} and Mrk~\num{348}; \citealt{Gallimore2001_ngc1068,Peck2003}); 
nuclear outflow masers trace the velocity and geometry of nuclear winds at a few parsecs from the nucleus (e.g., Circinus;~\citealt{Greenhill2003}).
All these studies are possible thanks to high angular resolution measurements made with the very long baselines interferometry (VLBI) technique.
Indeed, VLBI observations allow to determine the distribution of the maser emission and the absolute position of the maser spots at milliarcsecond (\si{\mas}) resolution and consequently to study the geometry of the nuclear regions. 
In particular, 
by analysing the disk masers, which are associated with the nuclear accretion disk, it is possible to estimate the rotation velocity and enclosed nuclear mass, namely the mass of the supermassive black hole (SMBH) \citep[e.g.,][]{Gao2017_3smbhmass, Pesce2020_geomdist}; to obtain distances to the host galaxies~\citep[e.g.,][]{Braatz2013IAUS,Reid2013} and to provide a direct estimate of the Hubble constant,  H$_0$ (\citealt{Reid2009, Pesce2020_H0}). 
The tracing of the accretion disk at sub-parsec scale permits to study the co-evolution of SMBH and galaxies in the lower-mass regime \citep{Greene2016} and to constrain the spin of obscured AGNs as recently proposed by \citet{Masini2022}.
The maser spectra associated with disks are identified by a three-peaked pattern, i.e. with three distinct groups of features: one around the systemic velocity of the galaxy and the other two at blue- and red-shifted velocities (in general, of the order of hundreds of \si{\kms} e.g., \citealt{Tarchi2012}). 
However, galaxies with only two maser line groups have been reported (e.g., Mrk\num{1210} Mrk1 and NGC\num{5728};~\citealt{Zhao2018, Kuo2020}).
Recently, \cite{Darling2017} discussed how maser radiation could  also be detected via gravitational lensing or deflection by massive black holes in inclined accretion disks (more than \num{10} degree from edge-on).
Based on this, \cite{Darling2017} suggests the existence of a new type of maser: the ``inclined maser disk'' and one of the candidates for this class is the galaxy \ic (see Sect.~\ref{sec:ic485}).
The observational signature of an inclined maser disk would be a narrow line, or lines complex, at the systemic velocity of the galaxy and at the apparent location  of the black hole. However, these peculiarities might also be due to other mechanisms and only VLBI observations can solve the ambiguity.

In order to determine the nature of the \water maser emission in the galaxy \ic, by providing the position and the distribution of the maser spots at mas resolution, we performed multi-frequency and multi-epoch Very Long Baseline Array (VLBA) and European VLBI Network  (EVN) observations. Their details are reported in Sect.~\ref{sec:obs} while the results are in Sect.~\ref{sec:results}. 
We discuss our results and we draw our conclusions in Sects.~\ref{sec:disc} and~\ref{sec:summ}, respectively.
We mention here that throughout the manuscript, we adopt a cosmology with $\Omega_{\text{matter}}= 0.27$, $\Omega_{\text{vacuum}}= 0.73$ and $H_0 = \SI{70}{\km \per \s \per \Mpc}$.
If not stated otherwise, the quoted velocities are calculated using the optical velocity definition in the heliocentric frame.

\section{IC$\,485$}
\label{sec:ic485}
IC\num{485} 
has been optically classified as an Sa spiral galaxy located at a distance of \SI{122.0(85)}{\Mpc} \citep{Kamali2017}. The spectroscopic classification of the galaxy is uncertain. \citet{Darling2017}  classify \ic as a LINER, while  \citet{Kamali2017} as a Seyfert \num{2}.
\cite{Darling2017} detected with the VLA a broad multi-component \water maser emission with a peak flux of around \SI{80}{\mJy}  that corresponds to a  isotropic luminosity of $L_{iso} = (\num{868(46)})\,L_{\text{\Sun}}$.
Furthermore, \cite{Darling2017} also reported the detection of an unresolved (with an angular resolution of around $\SI{90}{\mas} \simeq \SI{50}{pc}$) and faint continuum radio source of I$_{\nu} =\,$\SI{77(15)}{\uJy \per \beam} at \SI{20}{\GHz}. 
IC 485 was also part of the FIRST\footnote{http://sundog.stsci.edu/first/catalogs.html} ($S_{\text{peak}}= \SI{3.0}{\mJy}$) and NVSS (\SI{1.4}{\GHz} $=$ \SI{4.4}{\mJy}; \citealt{condon02}) surveys and although the presence of an AGN cannot be excluded, its dominant energy source is believed to be star formation \citep[see][and references therein]{Darling2017}.
At \SI{33}{\GHz}, a tentative (\num{3.5}$\sigma$) radio source of $~$\SI{66}{\uJy}  was reported by \citet{Kamali2017}.
Furthermore, the galaxy was sampled in  a high sensitivity single dish analysis in \cite{Pescegbt2015} 
and more recently,  in a survey with the purpose of observing the water maser transition at \SI{183}{\GHz} 
in \cite{Pesce183ghz2023}.


\section{Observations and data reduction}
\label{sec:obs}
In the following we describe the observations, the calibration, and the analysis of the data at three different bands acquired in six different VLBI epochs.
The main details of the observations are summarised in Table~\ref{tab:sintesi_osservazioni}.
We reduced and analysed the data of all epochs with the NRAO Astronomical Image Processing System (AIPS\footnote{http://www.aips.nrao.edu/}) software by following standard procedures.
A brief note about the naming of epochs employed by the authors: these are preceded by the array used to observe, so, for example, the epoch ``VLBA 2018.16'' refers to epoch 2018.16 observed with VLBA.
We introduce, here, our definition of ``maser component'': a group of one or more features which can be fit with a single Gaussian in a certain range of velocity. This will be adopted throughout the rest of the paper.


\begin {table*}
\caption []{Observational details.} 
\centering
\begin{tabular}{ l c c c c c c c c c c}
\hline
\hline
Array -  &  \multirow{2}{*}{$\nu$}      & Project   &  Observation      & Epoch      & Integr.    & Restoring              & Position & \multirow{2}{*}{rms\tablefootmark{a}} &  Participating \\ 
Band     &             &  code     &    date           &            & time       &  beam size               & Angle    &                      & stations\tablefootmark{c} \\ 
         & (\si{\GHz}) &           &                   & 2018.      & (hr)       & (mas~$\times$~mas)  & (\si{\degree})     & ($\frac{\rm{mJy}}{\rm{beam}}$) \\ 
\hline
\multirow{2}{*}{VLBA-L}   & \multirow{2}{*}{1.6}        &\multirow{2}{*}{BT\,142}  &\multirow{2}{*}{ 11 Feb. 2018}       & \multirow{2}{*}{11}       &  \multirow{2}{*}{2.5}    & \multirow{2}{*}{13.0 $\times$ 5.2}  &\multirow{2}{*}{-14}       &\multirow{2}{*}{0.029}     & Br, Fd, Hn, Kp, La,  \\   
         &             &          &                    &            &          &                     &             &              &  Mk, Nl, Ov, Pt, (Sc) \\
    \multirow{2}{*}{VLBA-K}   &  \multirow{2}{*}{22}        &\multirow{2}{*}{BT\,142}  & \multirow{2}{*}{26 Feb. 2018}       & \multirow{2}{*}{16}         &\multirow{2}{*}{2.5}    &  \multirow{2}{*}{1.2 $\times$ 1.0} &  \multirow{2}{*}{ 7.46}      & \textbf{3.02}\tablefootmark{b} & Br, Fd, (Hn), Kp, La, \\
         &             &          &                    &            &          &                     &             &     0.064          & Mk, Nl, Ov, Pt, (Sc)  \\
\multirow{2}{*}{EVN-C }  &\multirow{2}{*}{ 5.0}        & \multirow{2}{*}{ET\,038}   & \multirow{2}{*}{ 25 May 2018}       & \multirow{2}{*}{39}         &  \multirow{2}{*}{ 1.4}    & \multirow{2}{*}{ 5.3 $\times$ 3.5}     &  \multirow{2}{*}{ -3}            & \multirow{2}{*}{0.028}        &   Jb, (Wb), Ef, Mc, Nt, O8,  \\
         &             &          &                    &            &          &                     &             &              & Tr, Ys, Hh, Sv, Zc, Bd, (Ib) \\
\multirow{2}{*}{EVN-L}   & \multirow{2}{*}{ 1.7}        & \multirow{2}{*}{ET\,038}   &  \multirow{2}{*}{28 May 2018}       & \multirow{2}{*}{40}         &  \multirow{2}{*}{ 2.3 }   &  \multirow{2}{*}{17.2 $\times$ 9.1}  &   \multirow{2}{*}{-16 }        &  \multirow{2}{*}{0.018}      & Jb, (Wb), Ef, Mc, O8,         \\    
         &             &          &                    &            &          &                     &             &              &  Tr, Hh, Sv, Zc, Bd, Sr\\
\multirow{2}{*}{VLBA-L}  & \multirow{2}{*}{1.6}        & \multirow{2}{*}{BT\,145}   & \multirow{2}{*}{ 5 Oct. 2018}       & \multirow{2}{*}{59}         &  \multirow{2}{*}{ 2.5}    & \multirow{2}{*}{ 11.4 $\times$ 4.3}   &   16    &   0.05         &   Br, Fd, (Hn), Kp, La,  \\
         &             &          &                    &            &          &                     &             &              &  Mk, Nl, Ov, (Pt), Sc\\
\multirow{2}{*}{VLBA-K}  & \multirow{2}{*}{ 22 }        & \multirow{2}{*}{BT\,145}  & \multirow{2}{*}{30 Oct. 2018}      & \multirow{2}{*}{83}         &  \multirow{2}{*}{ 3.5 }   & \multirow{2}{*}{ 0.9 $\times$ 0.4}   &   \multirow{2}{*}{-10.91}   &  \textbf{5.87}\tablefootmark{b} &  Br, Fd, Hn, Kp, La,\\
         &             &          &                    &            &          &                     &             &     0.066         &   Mk, Nl, Ov, Pt, Sc  \\

\hline
\end{tabular}
\tablefoot{
\tablefoottext{a}{The rms has been estimated with the task IMEAN.}
\tablefoottext{b}{The noise in the cube map, indicated in boldface, is estimated in a range of 500 channels where there is not maser emission. The value reported is the average of the noise for each channels considered in the range.}
 \tablefoottext{c}{Station codes are as follows. VLBA, Br: Brewster; Fd: Fort Davis; Hn: Hancock; Kp: Kitt Peak; La: Los Alamos; Mk: Mauna Kea; Nl: North Liberty; Ov: Owens Valley; Pt: Pie Town; Sc: St. Croix. EVN, Jb: Jodrell Bank; Wb: Westerbork Synthesis Radio Telecope; Ef: Effelsberg; Mc: Medicina; Nt: Noto; O8: Onsala; Tr: Torun; Ys: Yebes; Hh: Hartebeesthoek; Sv: Svetloe; Zc: Zelenchukskaia; Bd: Badary; Sr: Sardinia Radio Telescope; Ib: Irbene. Telescopes in parentheses were scheduled but did not take part in the observations or did not produce good data due to technical problems.}
}
\label{tab:sintesi_osservazioni}
\end{table*}

\subsection{K Band -- Spectral line observations}
\subsubsection{VLBA observations and data reduction}
\label{sec:obs_spect_vlba}
We observed the $6_{16} \rightarrow 5_{23} \, \element{H_{2}O}$ maser transition toward \ic with eight and ten antennas of the VLBA on Febraury 26, 2018 (epoch 2018.16) and on October 30, 2018 (epoch 2018.83) under project codes BT142 and BT145 (P.I: A. Tarchi), respectively.
In epoch 2018.16 we observed with one IF of \SI{64}{\MHz} (\SI{870}{\kms})  centred at the systemic velocity of the galaxy ($V_{sys}$=\SI{8338}{\kms}), while in epoch 2018.83, we observed with two slightly-overlapped IFs of \SI{64}{\MHz} each, placed in such a way to span a total velocity range of $\sim$\SI{1400}{\kms}, which is enough to cover the systemic and the red-shifted maser components reported by \cite{Pescegbt2015}. 
The IFs of both epochs were correlated with 4096 spectral channels (resolution of \SI{15.6}{\kHz} corresponding to \SI{0.2}{\kms}) using the DiFX correlator \citep{Deller2011}.
 We observed in phase-referencing mode with cycle phase-reference - target of \SI{45}{\s} -- \SI{45}{\s} to correct the atmosphere phase variation and to measure absolute positions of the maser.   We used J0802+2509 ($\alpha_{J2000}=\,$08$\rm{^h}$ 02$\rm{^m}$ 41.58742257$\rm{^s}$;  $\delta_{J2000}=\,$+25\si{\degree} 09\arcmin 10\arcsec.8982794 \footnote{http://www.vlba.nrao.edu/astro/calib/}) at $\sim$\SI{1.7}{\degree} from the target  as a phase calibrator. The bandpass and the amplitude corrections  were made using the fringe-finders calibrators DA\num{193} and $4\text{C}39.25$. 
The total observing time was 6 and 8 hours in epochs 2018.16 and 2018.83, respectively.
 Then we self-calibrated the phase reference source \phaseref and the solutions obtained were transferred to the target.
Finally in both epochs, we produced, by using the AIPS task IMAGR, the total intensity image cube and a continuum map by averaging the line--free channels.
 In epoch 2018.83, the image cube was produced after gluing together the two IFs by using the AIPS task UVGLU. 

To identify the maser features, we analysed the map obtained by summing the channels that should contain the maser emission with the task SQASH. The channel range was selected considering the single-dish spectra reported in \citet{Pescegbt2015} and \citet{MCP}\footnote{https://safe.nrao.edu/wiki/bin/view/Main/MegamaserCosmologyProject}.  
We visualised this map and we identified bright spots by eye.
The spectra were then produced using the task ISPEC setting a window of 3 $\times$ 3 pixels (mainly covering the beam size and where 1 pixel = 0.1 mas) around the spots identified in the sqashed map.
The  absolute position was estimated by fitting the brightest maser spot of each maser feature with a two-dimensional Gaussian fit with the task JMFIT. We extracted the spectrum of each maser feature in an ASCII file through the task ISPEC. Such files were imported in the GILDAS software called CLASS\footnote{https://www.iram.fr/IRAMFR/GILDAS},  where we analysed the line properties (e.g., peak flux densities and peak velocity).
The signal-to-noise ratio (S$/$N) threshold value chosen for considering a maser detection real was $5 \sigma$.
However, we considered also any emission found in the velocity range of features observed and reported in the single dish spectrum \citep{Pescegbt2015,MCP}.
The continuum sources were identified in the continuum map considering only the emissions with S$/$N$\,\geq 5 \sigma$, that were Gaussian fitted with the AIPS task JMFIT.

\subsubsection{Absolute positions uncertainties}
\label{sec:obs_uncer-rel}

To measure the uncertainty of the absolute position (named $\Delta\alpha$ and $\Delta\delta$ for $\alpha_{J2000}$ and $\delta_{J2000}$, respectively) of the maser features and of the continuum sources in IC~\num{485},  we considered 
the  position errors of the phase calibrator\footnote{They are tabulated in NRAO's catalogue (\cite{calibratori} http://www.vlba.nrao.edu/astro/calib/)}  \phaseref  and the position errors due to the thermal noise ($\varepsilon_{\text{rms}}$) of the maps that is evaluated following Equation 13 of \citet{Reid2014} that we report here, for simplicity: $\varepsilon_{\text{rms}} \approx (0.5 \, \theta_{\text{beam}})/\text{SNR} $, where $\varepsilon_{\text{rms}}$ is the thermal noise, $\theta_{\text{beam}}$ is the dimension of the beam and the SNR is the signal-to-noise ratio.
These values are reported in Table~\ref{table:err_term}.

\begin{table*}
\caption{Uncertainties considered in evaluating the absolute position errors ($\Delta \alpha$ and $\Delta \delta$) of the VLBA K-band epochs.}  
\label{table:err_term}      
\centering                          
\begin{tabular}{c c c c c c c }        
\hline \hline                         
\multirow{2}{*}{Epoch}  &\multicolumn{2}{c}{Phase reference} &  $\varepsilon_{\text{rms}}$    & \multicolumn{2}{c}{$\Delta\alpha$} & $\Delta \delta$  \\
                          & $\varepsilon_{\alpha_{2000}}^{P.R.}$ & $\varepsilon_{\delta_{2000}}^{P.R.}$     &            &        &   & \\   
           & \multicolumn{2}{c}{(mas)}                       & (mas)   & (mas) &(${}^{s}$)  & (mas) \\
\hline
VLBA 2018.16 & \multirow{2}{*}{0.09} & \multirow{2}{*}{0.13} & 0.02 &  0.09   & \num{6e-6} & 0.14 \\ 
VLBA 2018.83 &                       &                       &0.05  &  0.10   & \num{7e-6}  &      0.14                \\
\hline
\end{tabular}
\tablefoot{The columns indicate the epoch considered; the position errors $\varepsilon_{\alpha_{2000}}^{P.R.}$ and $\varepsilon_{\delta_{2000}}^{P.R.}$, for the right ascension and declination of the phase reference (P.R.) \phaseref, respectively; the thermal noise computed according to \citet{Reid2014}, $\varepsilon_{\text{rms}}$; and the total uncertainly calculated for $\alpha_{J2000}$ and $\delta_{J2000}$, indicated as $\Delta\alpha$ and $\Delta\delta$, respectively.
}
\end{table*}
%
%
\subsubsection{Imaging of the data-sets}
To realise the continuum and cube maps, the data were Fourier-transformed using natural weighting and deconvolved using the CLEAN algorithm \citep{hoegbom74}. To map the continuum, we covered a field of 0.2 $\times$ 0.2 arcsecond$^2$ (corresponding to $\sim$ 120 $\times$ 120 pc$^2$), meanwhile to make the  cube map we mapped a field of 0.04 $\times$ 0.04  arcsecond$^2$ (corresponding to $\sim$ 24 $\times$ 24 pc$^2$). Both maps were centred at the position of the main maser feature, i.e. M1. 
Details of the maps produced are reported in Table~\ref{tab:sintesi_osservazioni}.

\subsection{L and C Band -- Continuum observations}


\subsubsection{VLBA observations}\label{sect:vlbaobs}
We observed \ic with the VLBA\footnote{The National Radio Astronomy Observatory is a facility of the National Science Foundation operated under cooperative agreement by Associated Universities, Inc.} of the NRAO  at L-band (1.6\,GHz) on February 11 and August 5, 2018, in two runs of 4.5 hours each (project codes BT142 and BT145 and P.I: A. Tarchi). The data were taken with four 64\,MHz IFs in dual circular polarisation and recorded at a rate of 2048 Mbps. The correlation of the data was performed using the DiFX software correlator \citep{Deller2011} with 128 channels per IF. 
Similarly to the spectral observations (see Sect.~\ref{sec:obs_spect_vlba}) we
 observed in phase-referencing mode,  (with cycles phase-calibrator - target of 1 min - 3 min, respectively), where the phase-calibrator is J0802+2509. 
 We also observed  as fringe finders and bandpass calibrators the strong compact quasars DA193 and 4C39.25 at the beginning and at the end of the observing run, in order to correct instrumental delays.

After some editing, where we flagged $\sim$23-30\% of the visibilities of IC~485, we applied the ionospheric corrections and the latest value of the Earth’s orientation parameters. Then, we corrected the delays and phases for the effect of diurnal feed rotation (parallactic angle) and the amplitudes for the digital sampler voltage offsets. We removed the instrumental delays caused by the passage of the signal through the electronics using the phase-calibration measurements associated with the data (in the PC table). At this point, we calibrated the bandpass using DA193 and 4C39.25, and the amplitude using the measured antenna gains and system temperatures. Finally, we self-calibrated the phase reference source J0802+2509 and the solutions were applied to the target source IC~485. 

\subsubsection{EVN observations}\label{sect:evnobs}
We observed the nucleus of \ic with the EVN\footnote{The European VLBI Network is a joint facility of independent European, African, Asian, and North American radio astronomy institutes.} at L- (central frequency $=\,$\SI{1.7}{\GHz}) and C-band (\SI{5.0}{\GHz}), in May 2018 (project codes: ET038; P.I: A. Tarchi). The data were recorded at 1024\,Mbps, with 8$\times$16\,MHz IFs and dual circular polarisation. The correlation of the data was performed using the EVN software correlator \citep[SFXC;][]{keimpema2015} at the Joint Institute for VLBI ERIC (JIVE), using 32 channels per IF.
We observed in phase-referencing mode  (with cycles phase-calibrator - target of  3 min - 7 min and  3 min - 5 min at L- and C-band, respectively)  to correct phase variation caused by the atmosphere  and to obtain information about absolute position.
The chosen phase calibrator  is the same of the VLBA observations, i.e. J0802+2509
and  the strong  compact sources J0854+2006 and DA193 were observed as fringe finders. 
The total observing time was 4 and 2.5 hours at L- and C-band, respectively.

Before calibrating, we inspected the data to look for radio frequency interference (RFI) and the so-called ``bad points'' or time ranges. We flagged $\sim$34-38\% of the visibilities of the target source at both bands. Differently from VLBA data, initial a-priori calibration, i.e. amplitude and parallactic angle calibration, was carried out by the standard EVN pipeline. We then calibrated the bandpass using all the fringe finders. Subsequently, we removed the instrumental delays by fringe fitting thee calibrators J0854+2006 and DA193. In order to solve for atmospheric phase variations, we self calibrated the phase reference source J0802+2509. Finally,  similarly to the VLBA calibration, we interpolated and applied the solutions of J0802+2509 to our target source IC~485. We followed the same calibration steps for both frequency bands, except that for the L-band dataset after the a-priori calibration we performed the ionospheric corrections as a first step of the calibration. 

\subsubsection{Imaging of the data-sets}
We used the same following procedure to image the EVN and VLBA data-sets. The data were Fourier-transformed using natural weighting and deconvolved using the CLEAN algorithm \citep{hoegbom74}. We mapped a field of 4$\times$4\,arcseconds$^2$ and 0.6$\times$0.6\,arcseconds$^2$ (corresponding to $\sim$2.4$\times$2.4\,kpc$^2$ and $\sim$360$\times$360\,pc$^2$) at L- and C-band respectively, centred at the position of the main maser feature, i.e. M1. In order to suppress the small scale sidelobs and increase the beamwidth, we have also produced a set of EVN maps down-weighting the data at the outer edge of the $(u, v)$ coverage, using a taper of 40 and 60\,M$\lambda$ at L- and C-band, respectively. Details of the best maps produced are reported in Table~\ref{tab:sintesi_osservazioni}. 

\section{Results}
\label{sec:results}
In this section, we report the outcomes from our data reduction.
The first sub-section is dedicated to K-band where we describe spectral and continuum results. Meanwhile, in the second sub-section, we expose the results of L- and C-band.

\subsection{K  Band}

\subsubsection{Maser}
\label{subsec:result_k-maser}
\begin{figure*}
\centering
\includegraphics[width = 16 cm]{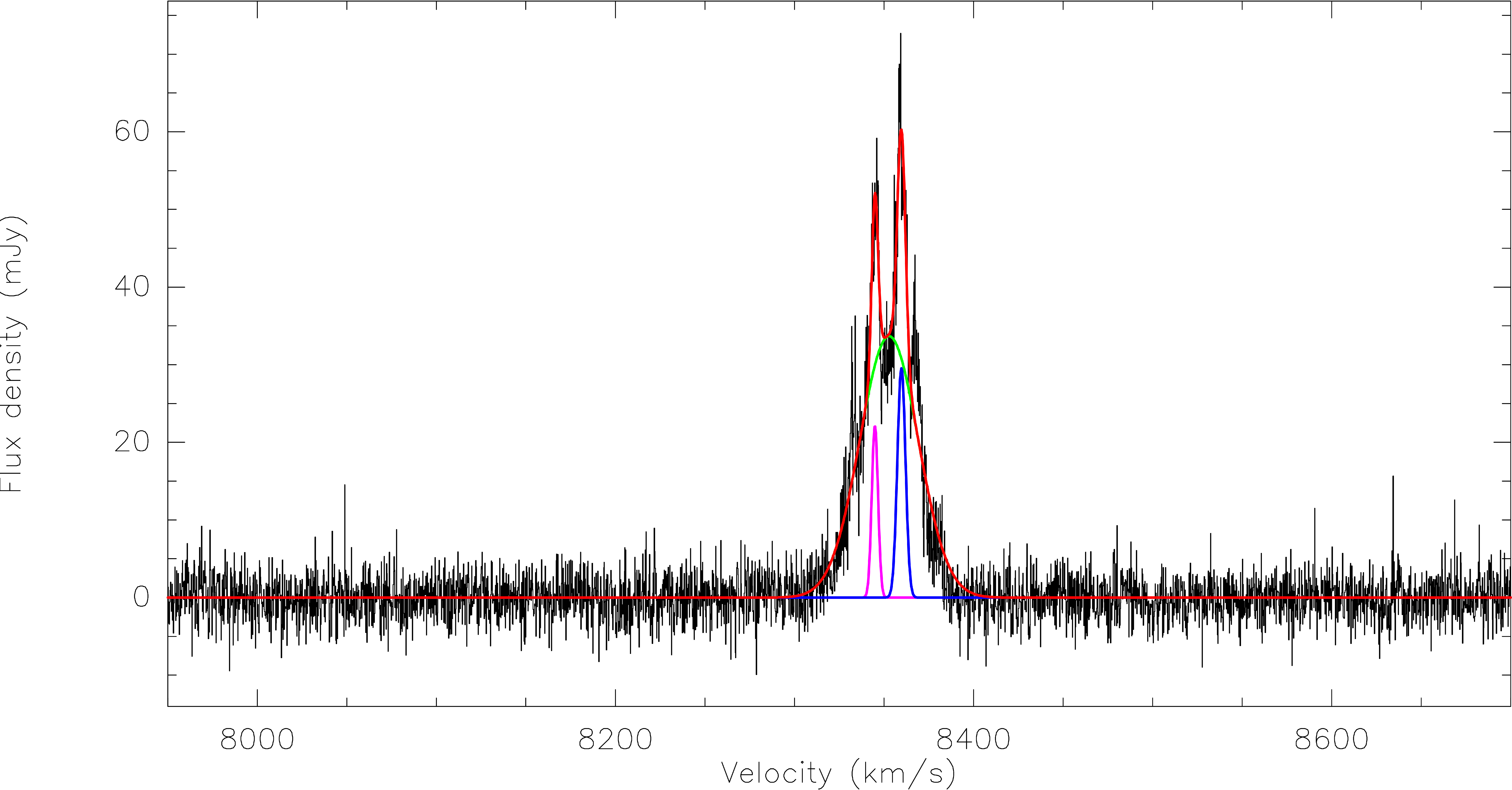}
\caption{Spectrum  of the component M\num{1} 
detected toward \ic with the VLBA in epoch 2018.16.
The thin red line is obtained by summing together the three Gaussian features. The three different Gaussian features are also reported in different colours (see Tab.~\ref{tab:maser}): in violet, green, and blue are feature 1, 2, and 3, respectively.}
\label{fig:maserM1_142}
\end{figure*}

\begin{figure*}
\centering
\includegraphics[width = 16 cm]{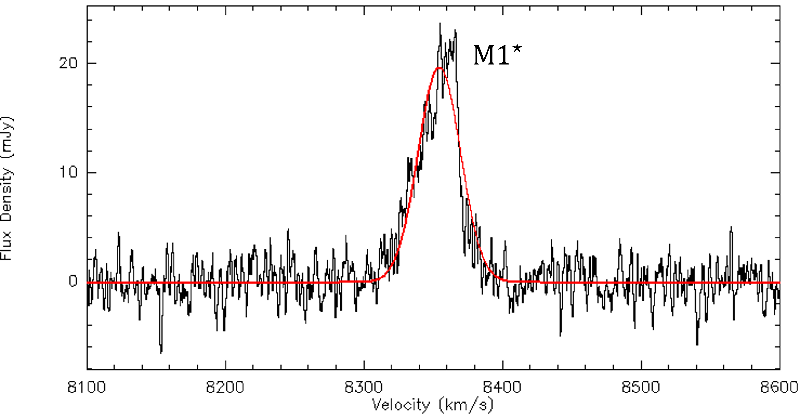}
\includegraphics[width = 16 cm]{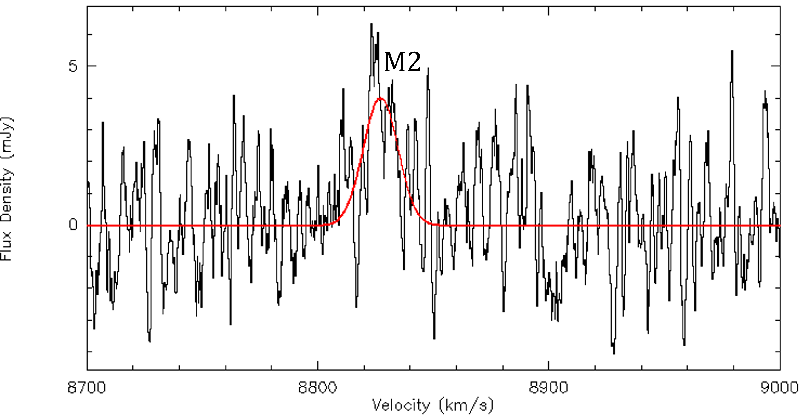}
\caption{ The spectra of the systemic component M1$^*$ (top panel) and of the red-shifted component (bottom panel) as detected in epoch VLBA 2018.83.
\textit{Top panel}: 
The spectrum covers the velocity range \SIrange{8100}{8600}{\km \per \s}. \textit{Bottom panel}: 
The spectrum covers the velocity range  \SIrange{8700}{9000}{\km \per \s}.
}
\label{fig:M1M2+fit}
\end{figure*}

In the two VLBA epochs  
we detected one \water maser component\footnote{We define ``maser component'' a group of one or more features which can be fit with a single Gaussian in a certain range of velocity.} close to the systemic velocity of \ic ($v_{sys}=\SI{8338}{\km \per \s}$) and one component shifted toward higher velocities. 
In particular,  the component at the systemic velocity has been observed in both epochs (named M1 and M\num{1}$^{*}$ in epochs 2018.16 and 2018.83, respectively) and the red-shifted component only in epoch 2018.83 (named M2). Their parameters are reported in Table~\ref{tab:maser}.
The component M1  has been fit using three Gaussian curves (see Fig.~\ref{fig:maserM1_142}):   
 a broad feature with a a full width at half maximum (FWHM) \SI{38.3}{\km \per \s} and two narrower Gaussian features with a FWHM of \num{4.1} and \SI{5.3}{\km \per \s}, respectively.
The estimated total isotropic luminosity of the three features is  \SI{526}{L_{\text{\Sun}}}.
The two components, M\num{1}$^{*}$ and M\num{2}, identified in VLBA 2018.83 are displayed in Fig.~\ref{fig:M1M2+fit}. 
The component M1$^{*}$ is composed of only one Gaussian feature and shows a flux density of \SI{19.7}{\mJy} and a linewidth  \SI{35}{\km \per \s}  that implies an isotropic luminosity of  \SI{239(28)}{L_{\text{\Sun}}}.
Also the component M2 is composed of one Gaussian feature and its flux density and line width are \SI{4}{\mJy} and \SI{18}{\km \per \s}, respectively,  with an isotropic luminosity of  \SI{24}{L_{\text{\Sun}}}.

\begin{table*}[t]
\caption{Parameters of the \SI{22}{\GHz} \element{H_{2}O} maser features detected in IC~\num{485}.}  
\label{tab:maser}
\centering
\small
\begin{tabular}{lccccccccc}
\hline
\hline
    & Maser         & R.A.                      & Dec.                      &    Gaussian     & Peak flux    &       FWHM         &      Velocity-integrated     & Peak velocity   &   $L_{iso}$ \\
    & component     &                           &                           &     feature     &  density    &                    &    flux density               &                 & \\
    &               & ($\rm{^{h}:~^{m}:~^{s}}$) & ($\rm{^{\circ}:\,':\,''}$)&                 & ( \si{\mJy} ) &  (\si{\km\per\s})  & (\si{\mJy \cdot \km \, \s^{-1}}) & (\si{\km\per\s}) & ($\,L_{\text{\Sun}}\,$) \\
         
\hline
\multirow{3}{*}{\rotatebox[]{90}{\tiny{2018.16}}}
& \multirow{3}{*}{M1} & \multirow{3}{*}{08:00:19.752530} &  \multirow{3}{*}{$+$26:42:05.0523} & 1 & \num{22(1)} & \num{4.1(2)}  & \num{97(4)} & \num{8344.98 (8)} & \multirow{3}{*}{\num{526(57)}\tablefootmark{a}} \\
&                     &                                  &                                & 2 & \num{33.7(5)} & \num{38.3(4)} &\num{1372(15)} & \num{8352.9(2)} & \\
&                      &                                &                                & 3 & \num{30(2)}  & \num{5.3(2)}  & \num{167(6)}& \num{8359.79(7)} & \\
\hline
\multirow{3}{*}{\rotatebox[]{90}{{\tiny{2018.83}}}}
&             &                &                 &  &              &             &          &     & \\
& M\num{1}$^*$  & 08:00:19.752516 & $+$26:42:05.0525 & 1 & \num{19.7(9)} & \num{35(1)} &  \num{742(21)} & \num{8354.8(5)} & \num{239(28)}  \\
&M\num{2} & 08:00:19.752516 & $+$26:42:05.0528 &  1  &  \num{4(1)} & \num{18(7)} & \num{76(20)} & $8827.0 \pm 1.0$ & \num{24(16)}      \\
\hline
\end{tabular}
\tablefoot{The first block summarise the parameters of the maser component of  epoch 2018.16, the second one those relating at epoch 2018.83 (both observed with VLBA).
The columns indicate the name of the maser component, right ascension and declination obtain with the task JMFIT in AIPS, peak flux density, the FWHM, the area, i.e the velocity-integrated flux density, and the peak velocity obtain from a Gaussian fit performed in CLASS. The isotropic luminosity derives from $L_{H_2O}/[L_{\text{\Sun}}]= 0.023 \times S dV/[\si{\Jy \km \per \s}] \times D^2/[\si{\Mpc}]$ where S is the peak flux, $dV$ is FWHM and $D = \SI{122.0(85)}{\Mpc}$, assuming $H_0 = \SI{70}{\km \per \s \per \Mpc}$~\citep{Kamali2017}.
\tablefoottext{a}{Estimated considering the sum of the three features.}
}
\end{table*}

\begin{figure*}
\centering
\includegraphics[width = 6 cm]{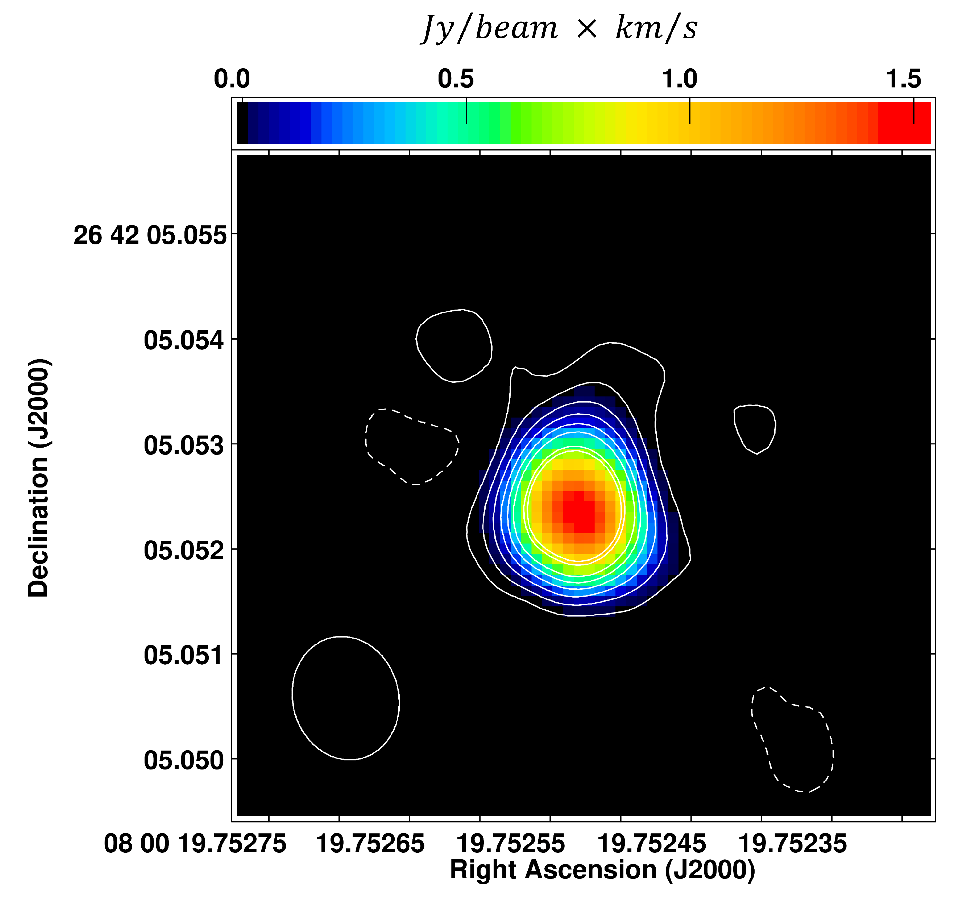}
\includegraphics[width = 6 cm]{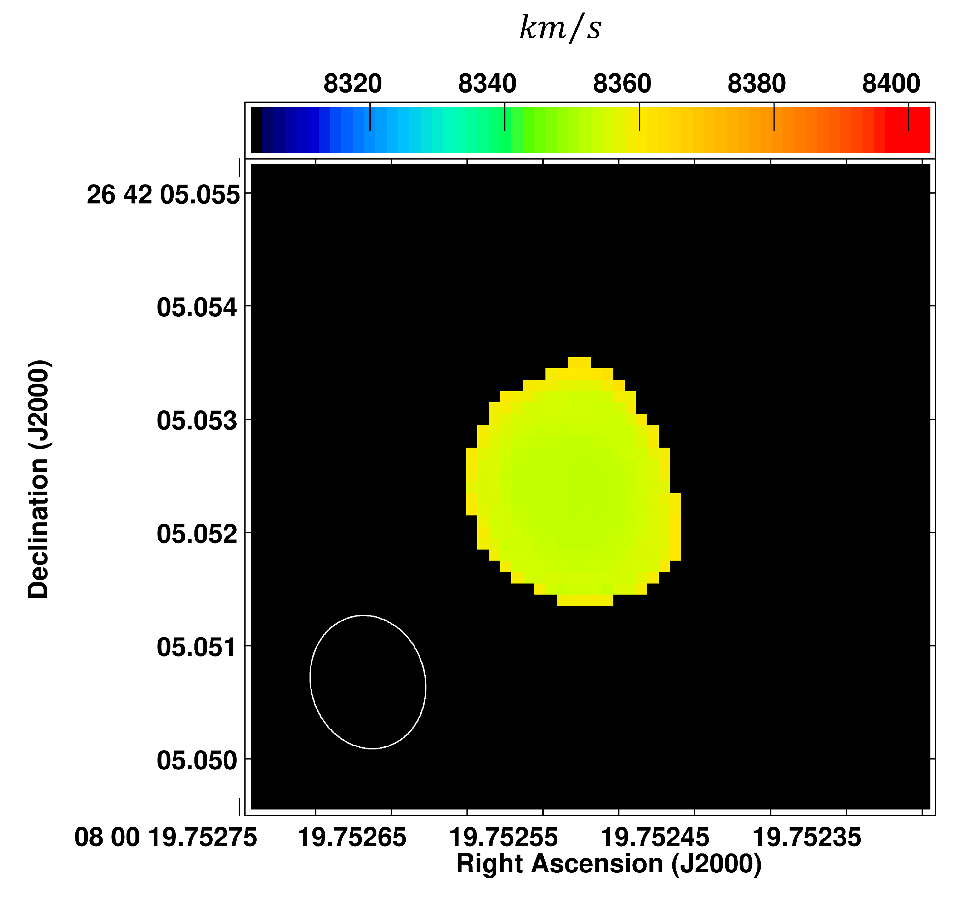}
\includegraphics[width = 6 cm]{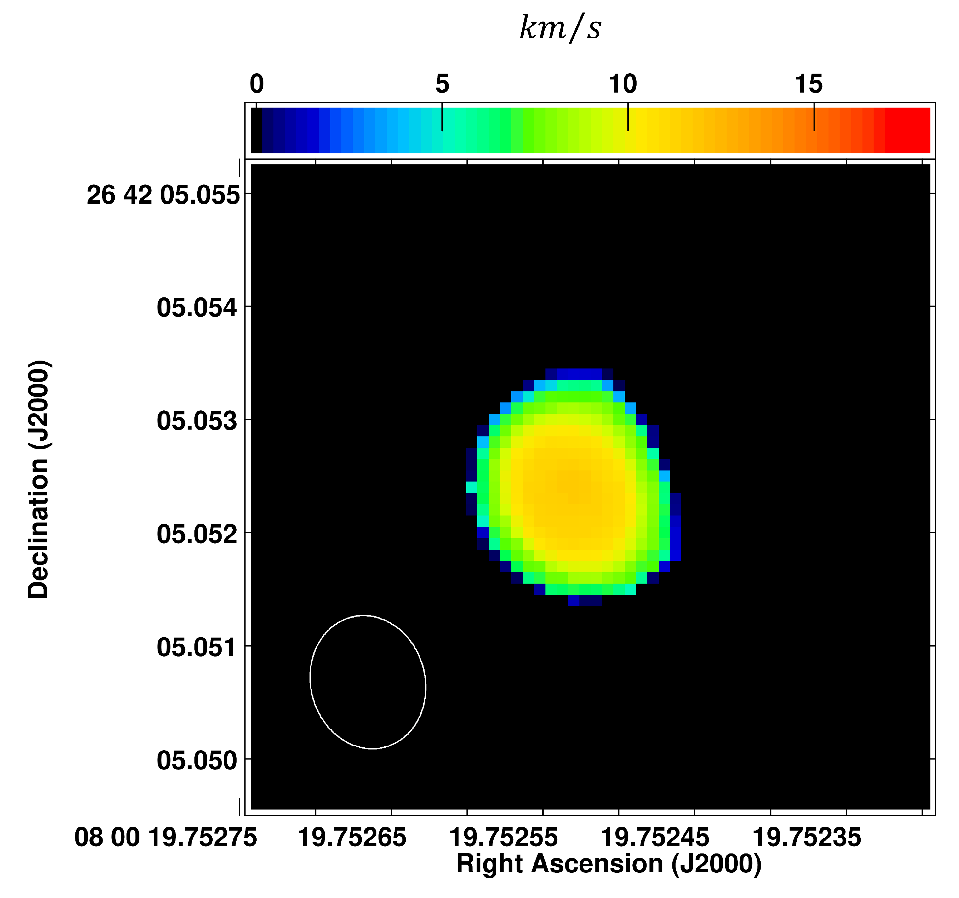}
\caption{Maps of the 0$^{\text{th}}$, 1$^{\text{st}}$, 2$^{\text{nd}}$ moment. \textit{Left panel}: Moment-zero map (colour scale) of the water maser emission in \ic superimposed on sqashed image (contours). Contour levels are: (-3, 3, 6, 9, 12, 15, 24, 26) $\times$ \SI{64}{\uJy \per \beam} ($1\sigma$ rms $=$ \SI{64}{\uJy \per \beam}).
\textit{Centre panel}: mean velocity (first moment) map. \textit{Right panel}: velocity dispersion (second moment) map. These figure are obtained from the epoch VLBA 2018.16.  In the bottom left corner of each  panel is reported the beam size (1.16 $\times$ 1.02) mas.
}
\label{fig:momzero}
\end{figure*}

\subsubsection{Coincidence between M1 and M1$^{*}$ }
\label{subsec:positionM1}
\begin{figure}
\resizebox{\hsize}{!}{\includegraphics{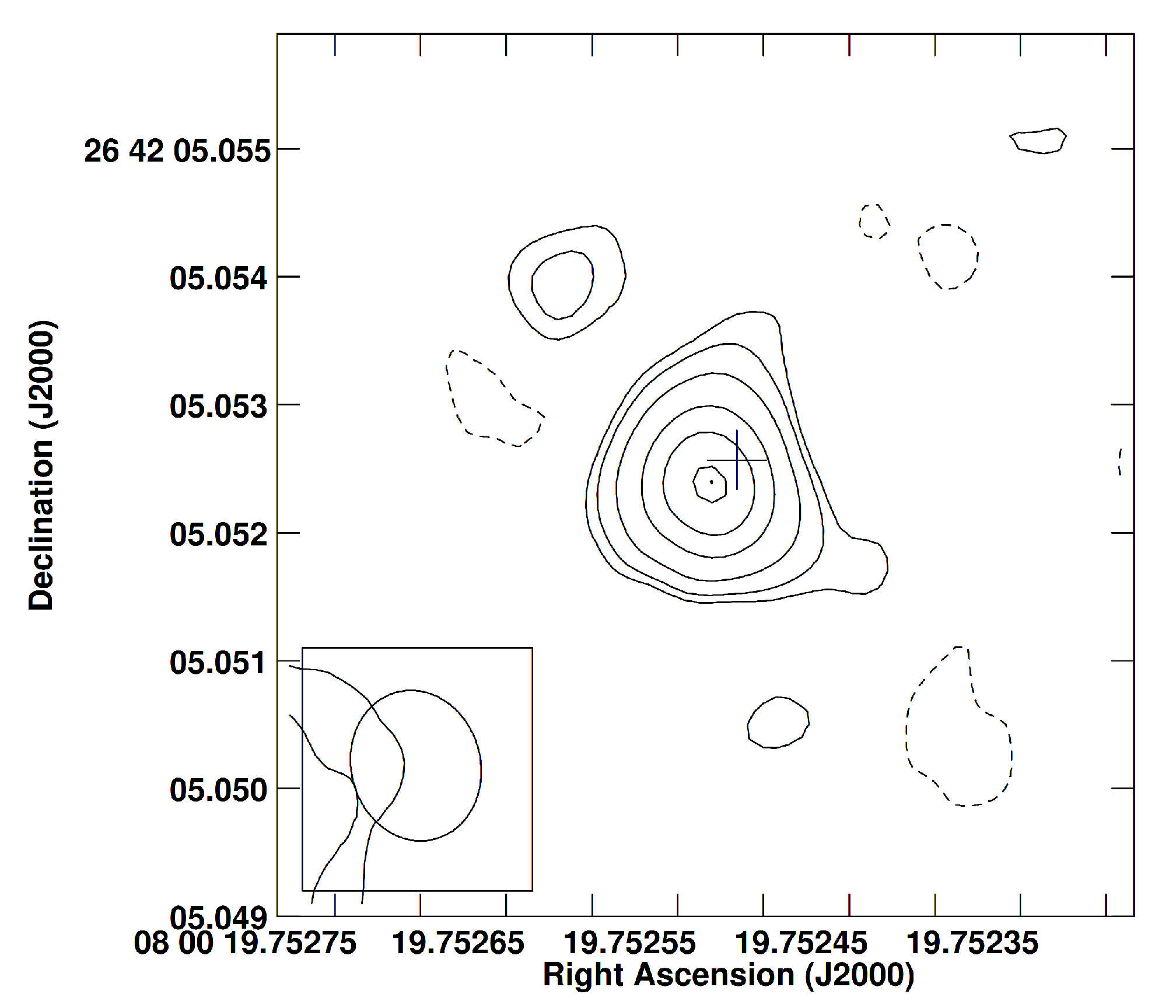}}
\caption{
Contours of the sqash map of the feature maser M1 in the epoch VLBA 2018.16. The cross indicated the position of the peak M1$^{*}$ in VLBA 2018.83. The $+$ symbol is proportional to relative error between the two epochs. 
 In the bottom left panel, the clean beam of VLBA 2018.16 is reported: ($0.78 \times 0.44$) mas.  
}
\label{fig:pos-relativa}
\end{figure}

We compared the absolute positions of M1 and M1$^{*}$, which are detected close to the systemic velocity of IC~485, in order to assess the coincidence of the two maser components. 
In particular, we summed the spectral channels within the line emission of M1 and M1$^{*}$, separately. Afterwards, we imaged their contour maps and by performing a 2D Gaussian fit, by using the AIPS task JMFIT, we estimated their absolute positions.
The results and the sqashed map are shown in Fig.~\ref{fig:pos-relativa}.
We also estimated the angular distance between M1 and M1$^{*}$, according the expression for small angular distance: $\vartheta \approx \sqrt{[(\alpha_{M1*} - \alpha_{M1}) \cos{\delta_{M1*}}]^2 + (\delta_{M1*} - \delta_{M1})^2}$; where $\alpha_{M1*} $, $\delta_{M1*} $ and $\alpha_{M1} $, $\delta_{M1} $ are the right ascension and the declination of M1$^{*}$ and M1, respectively. We obtained an angular distance of $\vartheta \approx (0.27 \pm 0.24) \,\si{mas}$.  
The error of the angular distance was calculated considered the values of Table~\ref{table:err_term}.
 As a matter of fact, the two positions are consistent within the error.
We conclude that the two components M1 and M1$^{*}$ are the same one, and hence, in the following we will refer to it as a single source, namely M1.

\subsubsection{Continuum}
\label{subsec:result_k-cont}

In  Table~\ref{tab:spot_continuo}, we report the absolute positions, the peak intensity, the integrated flux, and the S/N of the identified continuum sources at  K-band in the two VLBA epochs.
The sources are labelled with a "C" followed by a number in ascending order according to the right ascension. 
In the VLBA 2018.16 epoch, three continuum sources (from C1 to C3) were detected above the threshold (S$/$N $\geq 5 \sigma$).
Six sources (from C4 to C9) were instead detected in the VLBA 2018.83 epoch. 
All sources detected were unresolved and their positions compared with that of the maser emission are shown in Fig.~\ref{fig:continuo}. Moreover, no spatial correspondence among the sources was observed in the two epochs.
Furthermore, none of these nine sources coincides with the tentative ones detected at L- and C-bands (see Sect.~\ref{subsec:res_LCband}).

\begin{table*}
\caption{Parameters of the identified continuum sources at K-band with the VLBA.}             
\label{tab:spot_continuo}      
\centering                          
\begin{tabular}{c c c c c c c }        
\hline\hline
  & Name & R.A.      &   Dec      & Integrated Flux & S$/$N\\
     &  & ($\rm{^{h}:~^{m}:~^{s}}$) &($\rm{^{\circ}:\,':\,''}$)  &  (\si{\mJy}) &  \\    
\hline  
\multirow{3}{*}{\rotatebox[origin=c]{90}{\tiny{2018.16}}}
&C1 & 08: 00: 19.739974 & $+$26: 42: 05.1319  & $0.32 \pm 0.07$ & 5.1\\
&C2 & 08: 00: 19.741886 & $+$26: 42: 05.1896  & $0.33 \pm 0.07$ & 5.3\\
&C3 & 08: 00: 19.744134 & $+$26: 42: 05.0024  & $0.34 \pm 0.07$ & 5.4\\
\hline
\multirow{6}{*}{\rotatebox[origin=c]{90}{\tiny{2018.83}}}
&C4 & 08: 00: 19.743560 & $+$26: 42: 05.1308  & $0.37 \pm 0.07$ & 5.7\\
&C5 & 08: 00: 19.745021 & $+$26: 42: 05.0658  & $0.42 \pm 0.07$ & 6.5\\
&C6 & 08: 00: 19.745057 & $+$26: 42: 04.9444  & $0.38 \pm 0.07$ & 5.8\\
&C7 & 08: 00: 19.748663 & $+$26: 42: 05.1722  & $0.33 \pm 0.07$ & 5.1\\
&C8 & 08: 00: 19.751774 & $+$26: 42: 05.0719  & $0.33 \pm 0.07$ & 5.1\\
&C9 & 08: 00: 19.763236 & $+$26: 42: 04.9749  & $0.33 \pm 0.07$ & 5.0\\
\hline                                   
\end{tabular}
\tablefoot{
The columns indicate the name, the coordinates in epoch J\num{2000}, the integrated flux and signal-to-noise ratio for the continuum source in the continuum map. Each of these is unresolved, for this reason we report only the integrated flux equals to the peak flux density per beam.  We assume an  position error of 0.24~mas as described in Sect.~\ref{subsec:positionM1}. The errors are obtained by a Gaussian fit performed with the AIPS task JMFIT.
}
\end{table*}

\subsection{L and C Band}
\label{subsec:res_LCband}
No continuum source was detected above the 5$\sigma$ noise level neither at 1.4 nor at 5.0 GHz in a region of 100 mas ($\sim$600\,pc) radius from the maser position (see Sect.~\ref{subsec:result_k-maser}), although there are a number of tentative sources between 3 and 5$\sigma$. 
Nevertheless, it is worth mentioning that a tentative source is visible in the most sensitive EVN map at L-band, with a peak flux density of 68\,$\mu$Jy (3.8\,$\sigma$), whose position ($\alpha$=08$^{\rm h}$ 00$^{\rm m}$ 19$^{\rm s}$.7522; $\delta$=26\degr 42\arcmin 05\arcsec.051) coincides, within the errors, with the VLA source detected at 20\,GHz by \citet{Darling2017}. 
However, there is no hint of this source in the most sensitive VLBA image, not even at the 2$\sigma$ level. With the aim of confirming the presence of this feature we have combined the EVN and the first VLBA data-sets in the $(u,v)$ plane and, then, imaged the resulting data-set using natural weighting. The noise of the cleaned image is  \SI{24}{\uJy \per \beam}. Unfortunately, the source was not detected in the combined map. 
This suggests that also this source is an artefact. 

\begin{table*}
\caption{Details of the radio continuum observation of IC~485.}             
\label{table:obs_literature}      
\centering          
\begin{tabular}{ccccccccccc}     
\hline\hline       
\multirow{2}{*}{Array} & \multirow{2}{*}{$\nu$} &   Angular   &      Linear   &  Angular       &    Linear      & \multirow{2}{*}{$S_{\rm peak}$} & \multirow{2}{*}{L$_{\nu}$\tablefootmark{b}} & \multirow{2}{*}{rms} & \multirow{2}{*}{Ref.} \\
      &       & resolution  & resolution & LAS\tablefootmark{a} &      LAS       &                &           &     &       \\
      & (GHz) &  (arcsec)   &     (pc)      &  (arcsec)      &      (kpc)     &   (mJy/beam)   & (\SI{e27}{\ergs}) &($\mu$Jy/beam) & \\
\hline
VLA  & 1.4    & 45    & 27$\times 10^3$     & 370   & 600      & 4.4       & 78.1     & 450 &  1  \\
VLA  & 1.4    & 5     & 3$\times 10^3$      & 120   & 72       & 3.01      & 53.4     & 150 &  2 \\
VLA  & 20     & 0.08  & 48                  & 2.4   & 1.4      & 0.077     & 1.37     & 18 &  3  \\
VLA  & 33     & 0.6   & 180                 & 44    & 26       & $<$ 0.086 & $<$ 1.53 & 19 & 4 \\
EVN  & 1.7    & 0.02  & 12                  & 0.12  & 0.06     & $<$ 0.09  & $<$ 1.60 & 18 & 5\\
EVN  & 5.0    & 0.005 & 3                   & 0.04  & 0.02     & $<$ 0.14  & $<$ 2.48 & 28 & 5 \\
\hline                  
\end{tabular}
\tablebib{
(1)~\citet{condon02}; (2)~FIRST\tablefootmark{c}; (3)~\citet{Darling2017}; (4)~\citet{Kamali2017}; (5) this work.
}
\tablefoot{
\tablefoottext{a}{LAS = large angular scale.}
\tablefoottext{b}{
The value of isotropic luminosity L$_{\nu}$ was calculated according the equation \citep{Karttunen2017}: $\text{L}_{\nu} = 4\pi \,S_{\rm peak}\, D^2$, where D = 122 Mpc~\citep{Kamali2017}.
}
\tablefoottext{c}{http://sundog.stsci.edu/first/catalogs.html}
}

\end{table*}

%
%
\section{Discussion}
\label{sec:disc}

\subsection{The nature of maser emission in IC~485}
\label{sec:disc_natureMaser}

Water maser emission was reported by \citet{Darling2017} with the VLA array in A-configuration with a resolution of around \SI{100}{\mas}.
Our VLBI measurements found the bulk of the emission is coincident with the position reported by \citet{Darling2017}, improving the absolute positional accuracy (around 1 mas)  by two orders of magnitude.

The main water maser emission was analysed through the  moment maps shown in Fig.~\ref{fig:momzero}. 
 From the analysis of the maps, we note that the emission is concentrated in the systemic velocity range: \SIrange{8350}{8360}{\km \per \s} and the masing gas shows a compact and uniform  distribution in the region of emission. From the zeroth moment, we notice that the bulk of the emission coincides with the position of the maser feature M1. The analysis of the first moment does not reveal any velocity gradient.
 The maps were made considering the emission above $3 \sigma$. A second map was produced with a cutoff at $2 \sigma$, where the components M1B in epoch 2018.83 and M1D in epoch 2018.16 were observed. Given the low-significance level of these,  they are treated as tentative and discussed only in Appendix~\ref{apdx}.

\begin{figure}
\resizebox{\hsize}{!}{\includegraphics{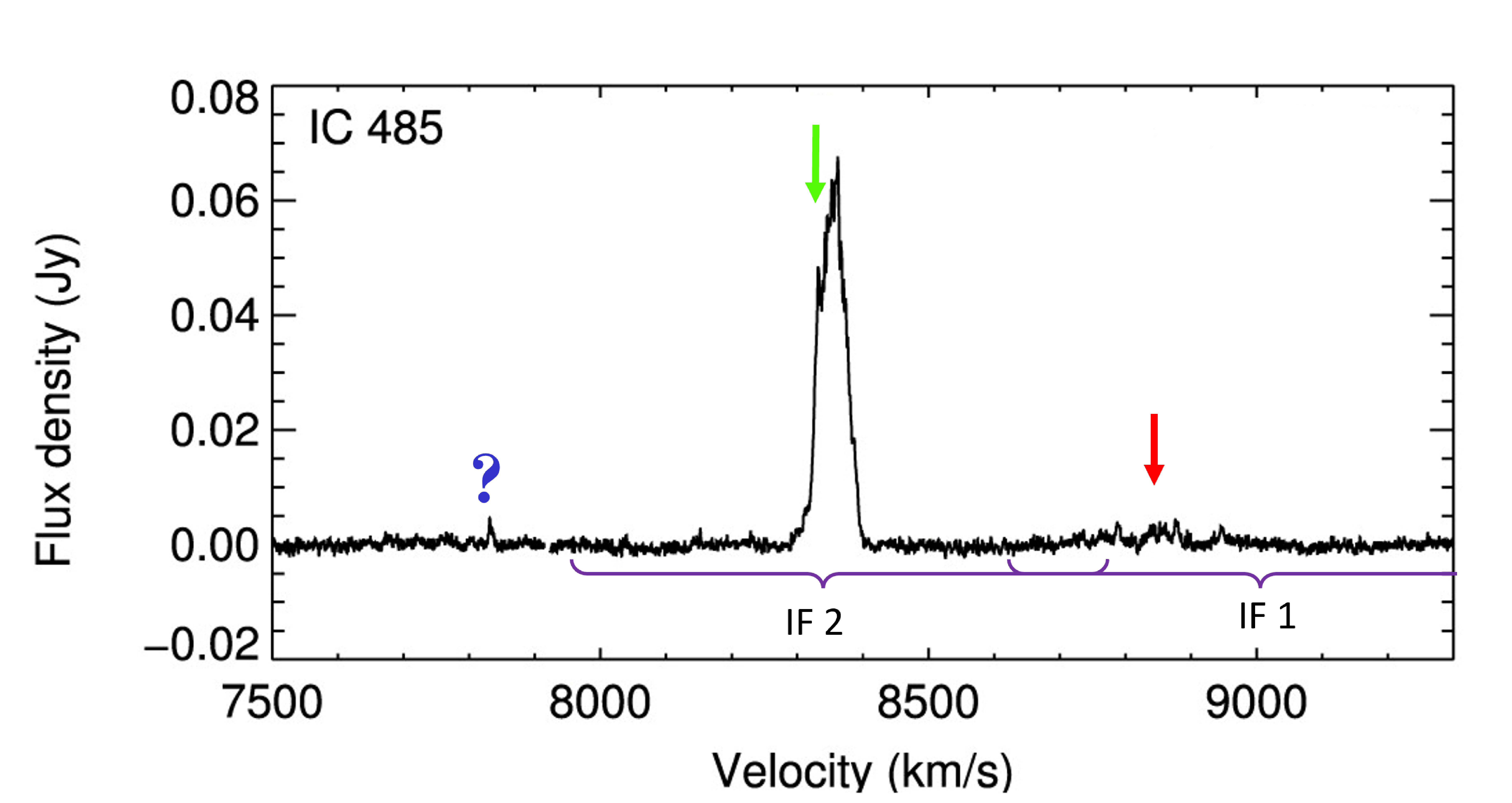}}
\caption{Spectrum of the water maser detected by \cite{Pescegbt2015} with the Green Bank Telescope (GBT) where we reported the velocity coverage (\SI{1400}{\kms}) of the two IFs used during the VLBA observations in epoch 2018.83.  An overlapping of about \SI{100}{\kms} is visible. The green and the red arrows  represent the velocities of the water maser components M1 and M2, respectively, as detected in the present work. The (tentative) blue-shifted component reported in  \cite{Pescegbt2015}  is highlighted with the blue question mark.}
\label{fig:PesceIFs}
\end{figure}

The detection of the red-shifted component M2, even though below $3\sigma$, is considered real. Indeed, its presence is strongly supported by the detection mentioned in \citet{Pescegbt2015}. Indeed, they reported a multi-epoch averaged and strongly sensitive spectrum (rms = \SI{0.74}{\mJy}) of IC~485. This spectrum shows the red- and blue-shifted components, though the latter is weak (\SIrange{2}{3}{\mJy}). 
In 2015, the red-shifted component showed a flux density of \SIrange{3}{4}{\mJy},  that is confirmed by our observations  (see Table~\ref{tab:maser}).
About the blue-shifted component, \citet{Pescegbt2015} reported it as tentative detection.
In the present work, it was not possible to observe it because of the bandwidth and the arrangement of the two IFs selected during the VLBA observations (see Sect.~\ref{sec:obs_spect_vlba}).  
Indeed, the IFs, whose number and size was dictated by the VLBA capabilities at the time of the observations, did not allow to cover the velocity range of the blue-shifted component (see Fig.~\ref{fig:PesceIFs}).

The presence of these features in the spectrum observed by \citet{Pescegbt2015}, 
the position of M\num{1}  at the systemic velocity and the red-shifted component M\num{2} observed in this work, together with their linear distribution lead us to support a disk nature of the maser. 
In addition, a recent survey made by \cite{Pesce183ghz2023} of \SI{183}{\GHz} \water maser emission from AGN known to host \SI{22}{\GHz} megamaser has shown that a significant fraction of the sample, including many known disk masers and among which IC\,485, hosts emission from both transitions. Additionally, some of the targets has triple-peaked spectra also at higher frequency. While in IC\,485, 183\,GHz emission is detected only close to the systemic velocity, likely due to the expected weakness of the satellite lines. The detection of the two water masers transition in \ic may still support and encourage our hypothesis of disk-maser. 

\begin{figure}
\resizebox{\hsize}{!}{\includegraphics{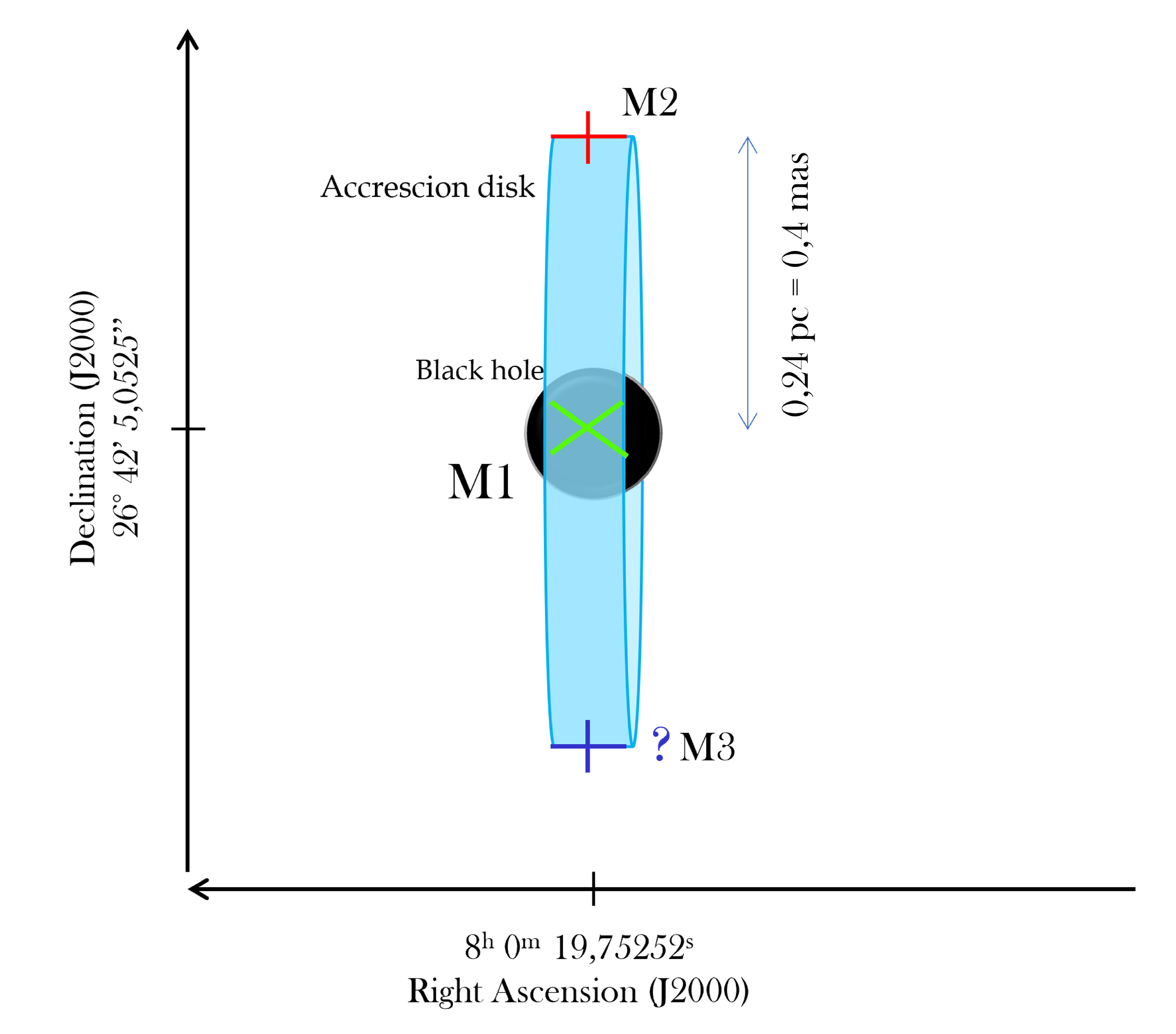}}
\caption{A portrayal of the disk geometry based on the water maser components. The disk (in cyan), that is assumed in Keplerian rotation, is edge-on and oriented north-south, with the black hole at the centre.
The positions of the detected maser components M1 and M2 are marked with a green ``$\times$'' and a red ``$+$'' symbols, respectively. The position of the supposed (the question mark reflects it) blue-shifted maser component M3 is indicated with a blue ``$+$'' symbol (see Sect.~\ref{sec:disc_natureMaser}).
}
\label{fig:modellino}
\end{figure}

Analysing the position of the systemic feature and that of the red-shifted components, we assume that they are two of the three typical components present in the full spectrum of a water maser associated with a disk. 
A simple portrayal of our idea is presented in Fig.~\ref{fig:modellino}. Here, we show the component M1 and M2, and we hypothesised the position of the blue-shifted component (called here M3) assuming that this is symmetrically opposite to M2. Furthermore, we also sketch the black hole and its accretion disk.
According to this scenario, the disk would be edge on with north-south orientation and rotating clockwise.  The angular dimension of the disk is  \SI{0.8}{\mas} that corresponds to a linear dimension of \SI{0.47}{\pc} at the distance of IC\,\num{485}. 
This value is consistent with that of the accretion disk in the galaxy NGC~\num{4258}~\citep[R$_{in} =\,$\SI{0.1}{\pc},][]{Herrnstein1997}, which is considered the prototype for the maser disks' studies, with that in  NGC~\num{1068} \citep[R$_{in} =\,$\SI{0.6}{\pc},][]{Morishima2023}, in Mrk~\num{1419} \citep[R$_{in} =\,$\SIrange{0.13}{0.43}{\pc},][]{Henkel2002}  and with the values measured in other galaxies hosting \water megamaser disks in which the radii are estimated in a range of \SIrange{0.03}{1.3}{\pc} \citep[e.g.,][]{Gao2017_3smbhmass,kuo2011} .
Assuming a Keplerian rotation, the relation $v^2 = GMR^{-1}$ gives the black hole mass at the centre of the nuclear region. According to this relation we have:

\begin{equation}
\centering
   \frac{M_{BH}}{M_{\text{\Sun}}} = 1.12 \times \left( \frac{v_r}{[\si{\km \per \s}]} \right)^2 \times \left( \frac{\theta}{[\si{mas}]} \right) \times \left(  \frac{D}{[\si{\Mpc}]} \right) \, , 
\end{equation}

where ${v_r}$ is the rotation velocity of the disk, that is the velocity difference between M1 and M2: ${v_r} \approx \SI{470}{\km \per \s}$; $\theta$  is the angular size of half of the disk (\SI{0.4}{\mas}; average value of the angular separation between M1 and M2 in the two epochs)  and $D$ is the distance of the galaxy ($D=\SI{122}{\Mpc}$).
Therefore, we found:
$$
M_{BH} = \num{1.2e7} M_{\text{\Sun}}  \,
$$
with systematic uncertainty dominated by the galaxy distance error.
The value is consistent with the one expected for a supermassive black hole in a LINER or Seyfert galaxy (e.g., \citealt{Kuo_2010}). 

\begin{figure*}
\centering
\includegraphics[width = 7 cm]{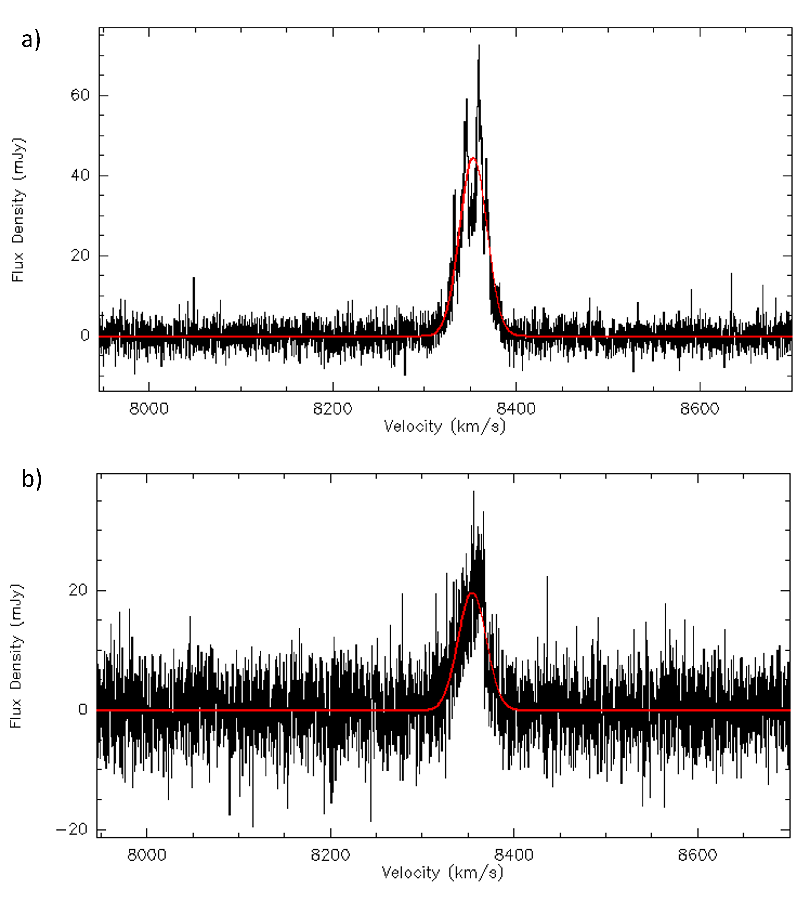}
\includegraphics[width = 11cm]{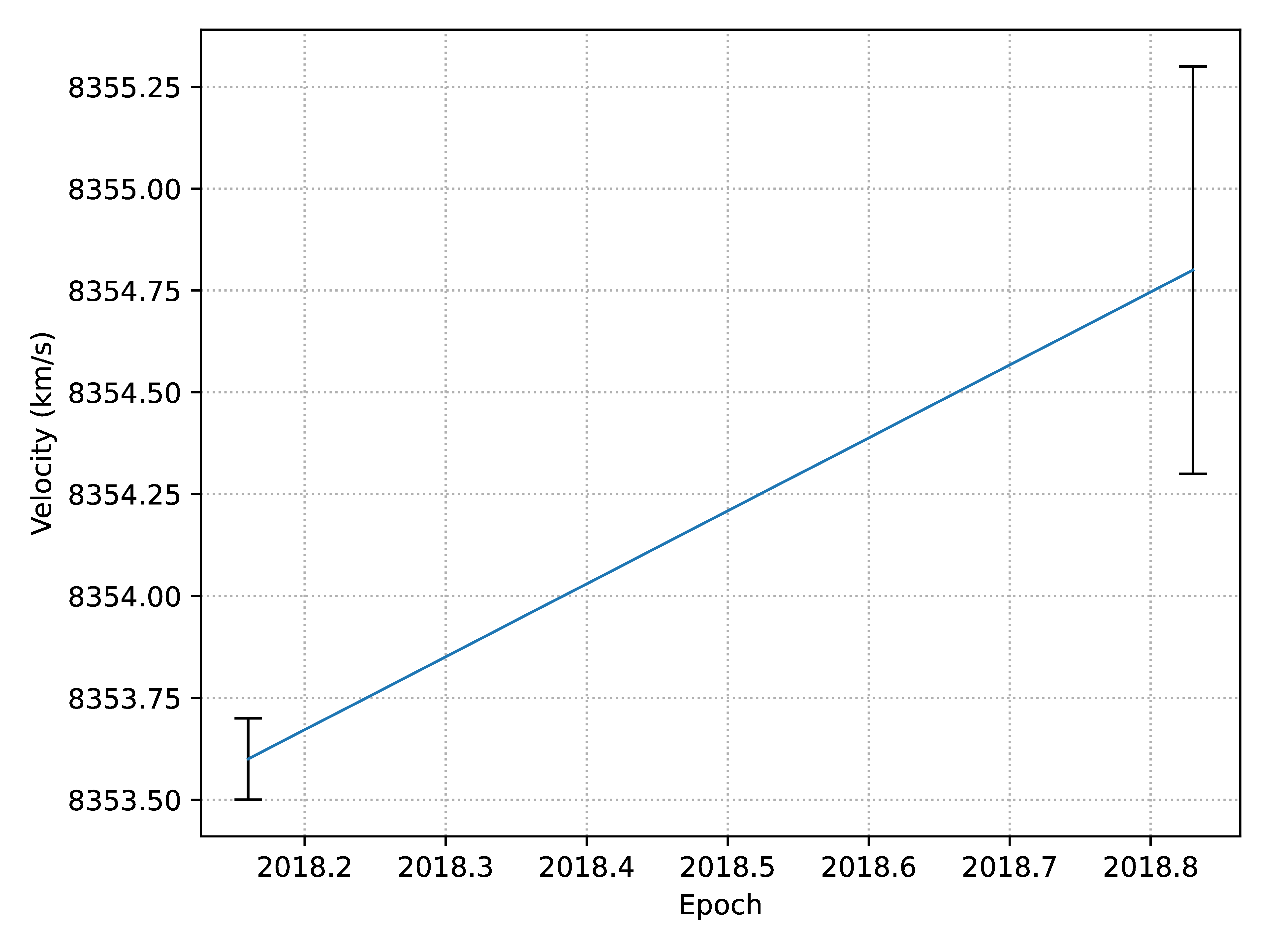}
\caption{Spectra and plots of velocity drift. \textit{Left panel}: The spectra  of the main maser component  M1 in  epoch 2018.16 (a) and in epoch 2018.83 (b) with the Gaussian fit (thin red line) used to estimate the velocity drift. \textit{Right panel}: The velocity-epoch diagram. In the diagram are reported the positions in velocity of M1 observed in the two epochs VLBA with the corresponding errors-bar derived by the Gaussian fit.
}
\label{fig:M1_4VD}
\end{figure*}

Assuming that the above scenario is correct, the detection of the main maser component, M1, in both K-band epochs, allows us to try a preliminary estimate of the velocity drift (usually this estimate is performed through a quasi-regular monthly-spaced monitoring program).
In order to determine this quantity, due to the lower flux density and less rich spectrum of the second epoch, we decided to fit the spectra, with no smoothing applied, of the maser component M1 with only one Gaussian feature  (see left panel of Fig.~\ref{fig:M1_4VD}). 
The velocity of the Gaussian peaks are  \SI{8353.6(1)}{\kms} and \SI{8354.8(5)}{\kms}, at epochs 2018.16 and 2018.83, respectively.
From a linear fit of these values (see right panel of Fig.~\ref{fig:M1_4VD}), we derived a velocity drift of $dv/dt \approx \SI{1.8}{\kmsy}$, with an uncertainty of $\pm \SI{0.6}{\kmsy}$ estimated from the error propagation.
Although  this value is roughly estimated, it is consistent with those observed in other maser disk (e.g., Mrk 1419 and NGC 6264; \citealt{Henkel2002,Kuo2013}).
This value can be also compared with the centripetal acceleration: $a_{c}=v{_r}^2/R$, derived by using the values obtained by us of the disk rotation velocity ($v_{r} \approx $ \SI{470}{\kms}) and of the disk radius ($R$ = \SI{0.24}{\pc}). The computed centripetal acceleration, $a_{c}=\SI{0.95}{\kmsy}\,$, is not fully consistent with that obtained for the velocity drift ($\,dv/dt=1.8 \, \pm \,0.6\, \si{\kmsy}$). In order to make our comparison more quantitative, however, we have to take into account the uncertainties in the parameters involved. While this is not trivial, given the impact of possible disk peculiarities (e.g., a slight deviation from perfectly edge-on orientation, warping, etc...), we can, anyway, associate an uncertainty to our estimate of the centripetal acceleration by assuming an error for R of \SI{0.09}{\pc}
(the positional error between M1 and M2) and of \SI{10}{\kms} for $v_r$ (based on the spread in velocity of the red-shifted features in the \citealt{Pesce2015}). This yields a value for the uncertainty of $\sim$ \SI{0.4}{\kmsy}. This computation somewhat reconciles the two values within the errors. However, these latter are likely to be even larger because of the aforementioned necessity to use a single Gaussian to fit entire M1 systemic feature complex impacts on our ability to  track the acceleration of each single feature (within a complex that extend over $\sim \SI{100}{\kms}$). The preliminary nature of our measurements further reinforces the necessity of an ad hoc monitoring program of the maser emission to better
constrain the velocity drift of the (systemic) maser lines. 
%

In addition, the maser component M1 in VLBA 2018.83 is characterised by a slow ascent and rapid descent (by observing the increasing axis of velocities) and defining an asymmetric shape of the line profile. This is expected for a rotating edge-on disk due to the mere effects of projection onto the line-of-sight of the emitting rays as explained in \cite{Schulz1995}. 
 Similar behaviour have been found in NGC~\num{4258}  and in many spectra of maser disks in the catalogue of \cite{MCP}.
The width of the maser line can be due to turbulent motions in the disk \citep[e.g.,][]{Pariev1998}
and speculatively attributable to a magneto-rotational instability as recently observed in  NGC~\num{4258} by \cite{Baan2022}.
Our hypothesis candidates \ic as new galaxy disk maser.
However, in order to confirm the disk nature, it is imperative to detect and to map the three main maser components at VLBI scales. To achieve this goal, the target should be further investigated through a high-sensitivity VLBI campaign.
Furthermore, observations with a high sensitivity array would also allow to clarify the nature of the tentative features observed. In particular, the tentative M1B and M1D presumably associated with jet/outflows maser which do not rule out a composite nature of the maser (see  Appendix~\ref{apdx}). Adding a new confirmed disk-maser at $\sim$100 Mpc would allow to exploit the source for black hole mass and distance estimates, pushing this kind of measurements farther out in the Universe, and strenghtening further the high potential impact of new upcoming facilities like the SKA and ngVLA with which we expect to lead similar studies up to high-z ($z \geq 1$).


%
%
%

%
%

\subsection{The origin of continuum emission: AGN and/or Star formation}
\begin{figure}
\resizebox{\hsize}{!}{\includegraphics[width=12cm]{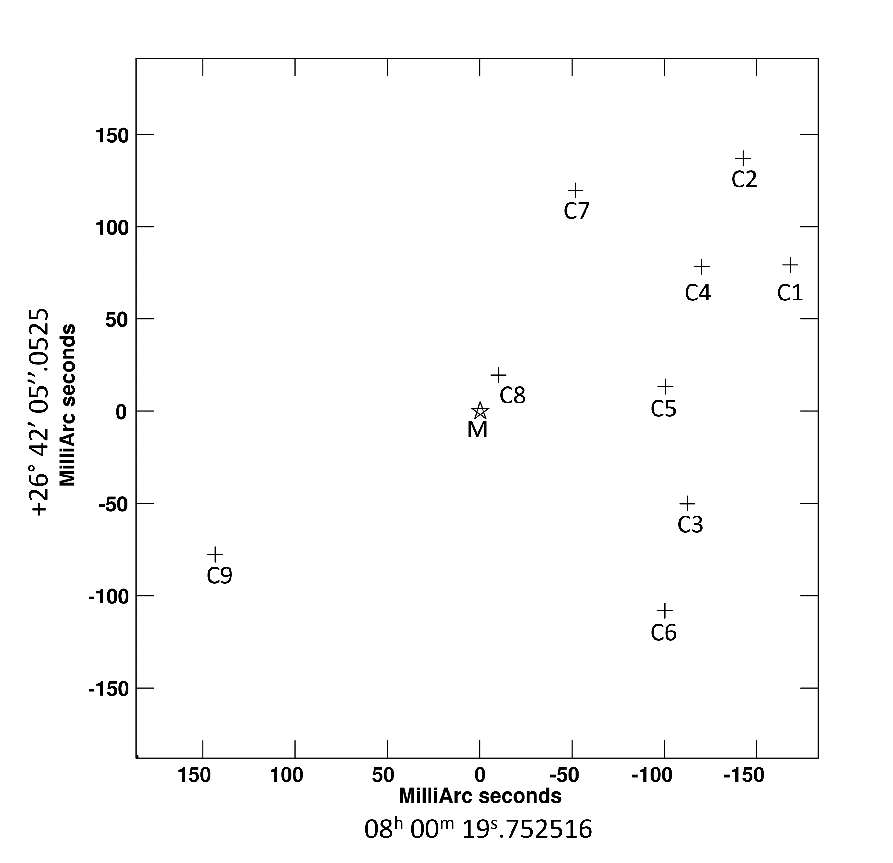}}
\caption{
Comparison of the absolute positions of the identified radio continuum compact sources (C1-9, see Table~\ref{tab:spot_continuo}) and of the water maser components M1 and M2 (see Table~\ref{tab:maser}). The ``$+$'' symbols indicate the compact sources while the star symbol is centred at the position of M1, (the dimension of the map do not permit to discern M1 to M2, for this reason, the point is indicated with ``M''). The size of the symbols do not correspond to the absolute positions uncertainties.
}
\label{fig:continuo}
\end{figure}

We identified nine radio continuum compact sources, all at K-band, around the water maser emission in the nuclear region of IC 485 (see Fig.~\ref{fig:continuo}).
The detection of these compact sources at only one of the three observed radio bands (no detections over $5\,\sigma$ at L- and C-band) and at pc scale can be explained in two ways: i) significant variability affecting the sources across the different epochs; ii) a highly-inverted spectrum for all the sources. A combination of the two factors may also be taking place.
From a statistical point of view, however, the possibility that such a large number of sources share the aforementioned characteristics (strong variability and/or inverted spectrum) is unlikely. Therefore, also considering the fact that, at K-band, all sources detected differ in position between the two epochs, we think that, especially for the weakest sources, we are actually dealing with spurious signal or artefacts.
However, this scenario leaves open the question on the absence of the radio emission expected to be present in the core of the AGN.


Indeed, as reported in Table~\ref{table:obs_literature}, unresolved and faint radio continuum emission was detected towards IC~485 at kpc-scales with the VLA at 1.4\,GHz, within the NVSS ($S_{\rm peak}$=4.4\,mJy; \citealt{condon02}) and FIRST\footnote{http://sundog.stsci.edu/first/catalogs.html} ($S_{\rm peak}$=3.01\,mJy) surveys. \citet{Darling2017} detected an unresolved radio compact  source in the nucleus of IC485 also at 20\,GHz, with the VLA, although with a barely sufficient significance ($S_{\rm peak}=77\pm15$\,$\mu$Jy, $\sim5$\,$\sigma$).

Considering the FIRST peak flux density of 3.01\,mJy, if we assume (as it was the case for IRAS\,15480-0534; \citealt{castangia2019}) that only the 30\% of the VLA flux is recovered in VLBI images, we would expect a flux density of about 0.9\,mJy in the L band map, well above our 5\,$\sigma$ noise level of 90\,$\mu$Jy. Therefore, the non-detection with the EVN  suggests that the kpc-scale radio emission observed with the VLA is mostly resolved out at pc scale, indicating a diffuse morphology. The bulk of the radio emission in IC~485 does not arise from a compact nuclear source, but it is diffused over a region larger than 0.1\,arcsec (60\,pc), the largest detectable angular scale of EVN observations. 
This suggests that the AGN in IC~485 is either radio silent (i.\,e., radio emission is entirely produced by star formation) or the AGN emits in the radio band but its emission (which might be produced by a jet, a nuclear wind, or a corona) is faint and the large (hundreds-parsec or kpc-scale) radio emission is dominated by the star forming regions in the host galaxy.

In order to estimate the expected radio emission from the AGN in IC~485, we can take advantage of the fundamental planes of black hole activity that link the radio luminosity of AGNs with their X-ray/optical luminosity and the black hole mass \citep[e.\,g.,][]{merloni03,baldi2021}. The standard fundamental plane is a correlation between the radio luminosity at 5 GHz, the 2--10\,keV X-ray luminosity, and the black hole mass (e.\,g., \citealt{merloni03}). 
\citet{Kamali2017} reported a Swift/BAT hard X-ray flux of $1.75\times10^{-11}$\,\si{\ergs}\,cm$^{-2}$ (corresponding to a luminosity of $\sim3.5 \times 10^{43}$erg\,s$^{-1}$), however, 
 we note that the large BAT PSF makes it difficult to distinguish if the emission is arising from \ic or, much more probably, from its companion IC~486, located at $\sim 5'$ from the former. \ic was observed also with Swift/XRT and XMM and a preliminary reduction of the latter data shows that is only marginally detected with an observed flux density of the order of $10^{14} \si{\ergs}\, \si{cm}^{-2}$ (corresponding to a luminosity of  $ \sim 10^{40} \si{\ergs}$ (L. Bassani privat. comm.).

Given the uncertainty in extrapolating a 2--10\,keV flux, therefore, we prefer to use the fundamental plane of black hole activity in the optical band found for the LeMMINGs sample by \citet{baldi2021}. We derive a rough estimate of the expected 1.5\,GHz core luminosity, by using the expression reported in \citet[][and reference therein]{baldi2021}:
\begin{equation}
    \centering
    \log L_{\rm core}=(0.83\log L_{[\rm O \sc III]}+0.82\log M_{\rm BH})\,m+q
\end{equation}

where $L_{\rm core}$ is the core luminosity, $L_{[\rm O \sc III]}$ is the [O$\rm \sc III$] emission line luminosity in units of erg\,s$^{-1}$ and $M_{\rm BH}$ is the black hole mass in units of M$_{\odot}$. The slope $m$ and the intercept $q$ differs for different type of galaxies (Seyferts, radio quiet LINERs, radio loud AGN and non jetted H$\rm \sc II$ galaxies; \citet[Table 2]{baldi2021}). Employing the black hole mass estimated through our maser study, $M_{\rm BH}\sim$1.2$\times10^7$\,M$_{\odot}$ (Sect.~\ref{sec:disc_natureMaser}) and $L_{[\rm O \sc III]}\sim 10^{40}$erg\,s$^{-1}$ (derived form the flux density reported in \citealt{zhu2011}), we obtain a core luminosity in the range $1\times10^{36} - 5\times10^{37}$\,erg\,s$^{-1}$. This luminosity range is consistent with the upper limit obtained from our EVN L-band observations ($3.1\times10^{36}$\,erg\,s$^{-1}$), suggesting that a more sensitive VLBI observations or an array with intermediate resolution between those of VLA and VLBI may potentially detect the weak radio emission expected from this AGN. 

%

Using the [OIII] luminosity, corrected for extinction as in \cite{Bassani1999}, we can estimate the intrinsic X-ray luminosity in the 2--10 \si{\keV} band from the correlation between L$_{\textit{X}}$ and L$_{\text{[OIII]}}$ found by \cite{Panssa2006}. Considering the observed L$_{\text{[OIII]}}$ and Balmer decrement, H$_{\alpha}$/H$_{\beta}$, reported in \cite{zhu2011}, we obtain $\text{L}_{\textit{X}} \sim 5 \times 10^{42} \si{\ergs}$. This luminosity is consistent with the observed flux in the case of a strongly absorbed or a low luminosity AGN.
The low radio luminosity observed in \ic can be explained by  investigating other properties and phenomenology of the host galaxy which affect it. Among these there may be SMBH mass and the accretion rate of material onto the black hole in conjunction with the inefficient accretion system \citep[e.g.,][and references therein]{Kamali2019}.
%
To clarify the spectroscopic classification of \ic, we cite new studies about AGNs and the distinction among various types of the latter, reported in \citet{Hechman2014}. These indicate how the difference of luminosity may be due to difference efficiency conversion of the potential energy of the gas accreted by the SMBH. In particular, the authors divide AGNs into two main categories: radiative-mode AGNs, historically called Seyfert galaxies and jet-mode AGNs, associated with LINERs galaxies. The low X-ray luminosity that characterise LINER galaxies would be due to the absence or the truncated geometrically-thin accretion disk and replaced by a geometrically-thick structure.
Generally, LINERs and low-luminosity AGNs show a X-ray luminosity in the range \num{e39}-- \SI{e41}{\ergs} \citep[e.g.,][]{Awaki1999,Terashima2002}. For this reason, the high hard-X luminosity observed in \ic  seems to point toward a classification as Seyfert (type 2).
Additionally, the possible presence and/or the dimensions of the inner radius of the accretion/maser disk of the target ($\sim \SI{0.24}{\pc}$, see Sect.~\ref{sec:disc_natureMaser}) seems to rule out the idea of a  truncated accretion disk and addresses the above classification (Sy2).
Further analysis is therefore needed, not only to confirm but also to study the actual size of the eventual disk.

\section{Summary}
\label{sec:summ}
In  this paper, we report a multi-epoch and multi-band radio VLBI study of the galaxy IC\,485. 
The observations were conducted in continuum and spectral modes using the VLBA and the EVN arrays. The outcome of our work allows us to obtain, for the first time, at mas scales, the distribution (and absolute position) of the maser and continuum emissions in the nucleus of the galaxy.

We have detected nine weak radio compact sources at K-band, but none at L- and C-bands. This indicates that at L- and C- bands the nuclear radio continuum emission reported at larger scales is diffused and it is resolved-out at our very high resolution. Furthermore, the relative weakness of the radio emission  is suggestive of a nuclear region hosting a radio silent AGN and/or dominated by regions of star formation. 

The maser emission has been detected in our VLBI maps at \SI{22}{\GHz} and it shows two spatially-distinguished linearly-distributed components: the first one at the systemic velocity (\SI{8355}{\kms}) and
the second one at a red-shifted velocity, offset by \SI{472}{\km \per \s} from the systemic velocity.
The scenario offered by our analysis supports a nature for the maser associated with an edge-on  accretion disk, with  north-south orientation and a radius of \SI{0.24}{pc}.
By assuming a Keplerian rotation, the estimated enclosed mass is of $M_{BH} = \num{1.2e7} M_{\text{\Sun}}$, consistent with those estimates in other galaxies belonging to the same AGN class (LINERs/Seyfert 2s) of IC\,485. 
The tentative detection of some additional maser features, displaced from the putative disk, may hint to a composite nature of the maser (part associated with the disk and part of jet/outflow origin) and deserve further investigation.

\begin{acknowledgements}
We thank the anonymous referee for useful comments on the manuscript. We are also grateful to Loredana Bassani for sharing information on high-energy data of IC485. E.L. and A.T. would like to thank Liz Humphreys for the constructive discussion during the `VLBI-40' meeting in Bologna.

\end{acknowledgements}

\bibliographystyle{aa} 
\bibliography{Bibliografia_articolo} 

\begin{thebibliography}{50}
\expandafter\ifx\csname natexlab\endcsname\relax\def\natexlab#1{#1}\fi

\bibitem[{Awaki(1999)}]{Awaki1999}
Awaki, H. 1999, Advances in Space Research, 23, 837, the AGN/Normal Galaxy

\bibitem[{{Baan} {et~al.}(2022){Baan}, {An}, {Henkel}, {Imai}, {Kostenko}, \& {Sobolev}}]{Baan2022}
{Baan}, W.~A., {An}, T., {Henkel}, C., {et~al.} 2022, Nature Astronomy, 6, 976

\bibitem[{{Baldi} {et~al.}(2021){Baldi}, {Williams}, {Beswick}, {McHardy}, {Dullo}, {Knapen}, {Zanisi}, {Argo}, {Aalto}, {Alberdi}, {Baan}, {Bendo}, {Fenech}, {Green}, {Kl{\"o}ckner}, {K{\"o}rding}, {Maccarone}, {Marcaide}, {Mutie}, {Panessa}, {P{\'e}rez-Torres}, {Romero-Ca{\~n}izales}, {Saikia}, {Saikia}, {Shankar}, {Spencer}, {Stevens}, {Uttley}, {Brinks}, {Corbel}, {Mart{\'\i}-Vidal}, {Mundell}, {Pahari}, \& {Ward}}]{baldi2021}
{Baldi}, R.~D., {Williams}, D.~R.~A., {Beswick}, R.~J., {et~al.} 2021, \mnras, 508, 2019

\bibitem[{{Bassani} {et~al.}(1999){Bassani}, {Dadina}, {Maiolino}, {Salvati}, {Risaliti}, {Della Ceca}, {Matt}, \& {Zamorani}}]{Bassani1999}
{Bassani}, L., {Dadina}, M., {Maiolino}, R., {et~al.} 1999, \apjs, 121, 473

\bibitem[{{Braatz} {et~al.}(2020){Braatz}, {Condon}, {Constantin}, {Gao}, {Hao}, {Henkel}, {Impellizzeri}, {Kuo}, {Litzinger}, {Lo}, {Pesce}, {Reid}, {Wagner}, \& {Zhao}}]{MCP}
{Braatz}, J., {Condon}, J., {Constantin}, A., {et~al.} 2020, Megamaser Cosmology Project

\bibitem[{{Braatz} {et~al.}(2018){Braatz}, {Condon}, {Henkel}, {Greene}, {Lo}, {Reid}, {Pesce}, {Gao}, {Impellizzeri}, {Kuo}, {Zhao}, {Constantin}, {Hao}, \& {Litzinger}}]{braatz2018IAUS}
{Braatz}, J., {Condon}, J., {Henkel}, C., {et~al.} 2018, in Astrophysical Masers: Unlocking the Mysteries of the Universe, ed. A.~{Tarchi}, M.~J. {Reid}, \& P.~{Castangia}, Vol. 336, 86--91

\bibitem[{{Braatz} {et~al.}(2013){Braatz}, {Reid}, {Kuo}, {Impellizzeri}, {Condon}, {Henkel}, {Lo}, {Greene}, {Gao}, \& {Zhao}}]{Braatz2013IAUS}
{Braatz}, J., {Reid}, M., {Kuo}, C.-Y., {et~al.} 2013, in Advancing the Physics of Cosmic Distances, ed. R.~{de Grijs}, Vol. 289, 255--261

\bibitem[{{Castangia} {et~al.}(2019){Castangia}, {Surcis}, {Tarchi}, {Caccianiga}, {Severgnini}, \& {Della Ceca}}]{castangia2019}
{Castangia}, P., {Surcis}, G., {Tarchi}, A., {et~al.} 2019, \aap, 629, A25

\bibitem[{{Cheung} {et~al.}(1969){Cheung}, {Rank}, {Townes}, {Thornton}, \& {Welch}}]{Cheung1969_1detezMaser}
{Cheung}, A.~C., {Rank}, D.~M., {Townes}, C.~H., {Thornton}, D.~D., \& {Welch}, W.~J. 1969, \nat, 221, 626

\bibitem[{{Condon} {et~al.}(2002){Condon}, {Cotton}, \& {Broderick}}]{condon02}
{Condon}, J.~J., {Cotton}, W.~D., \& {Broderick}, J.~J. 2002, \aj, 124, 675

\bibitem[{{Darling}(2017)}]{Darling2017}
{Darling}, J. 2017, \apj, 837, 100

\bibitem[{{Deller} {et~al.}(2011){Deller}, {Brisken}, {Phillips}, {Morgan}, {Alef}, {Cappallo}, {Middelberg}, {Romney}, {Rottmann}, {Tingay}, \& {Wayth}}]{Deller2011}
{Deller}, A.~T., {Brisken}, W.~F., {Phillips}, C.~J., {et~al.} 2011, \pasp, 123, 275

\bibitem[{{Gallimore} {et~al.}(1996){Gallimore}, {Baum}, \& {O'Dea}}]{Gallimore1996}
{Gallimore}, J.~F., {Baum}, S.~A., \& {O'Dea}, C.~P. 1996, in American Astronomical Society Meeting Abstracts, Vol. 189, American Astronomical Society Meeting Abstracts, 109.05

\bibitem[{{Gallimore} {et~al.}(2001){Gallimore}, {Henkel}, {Baum}, {Glass}, {Claussen}, {Prieto}, \& {Von Kap-herr}}]{Gallimore2001_ngc1068}
{Gallimore}, J.~F., {Henkel}, C., {Baum}, S.~A., {et~al.} 2001, \apj, 556, 694

\bibitem[{{Gao} {et~al.}(2017){Gao}, {Braatz}, {Reid}, {Condon}, {Greene}, {Henkel}, {Impellizzeri}, {Lo}, {Kuo}, {Pesce}, {Wagner}, \& {Zhao}}]{Gao2017_3smbhmass}
{Gao}, F., {Braatz}, J.~A., {Reid}, M.~J., {et~al.} 2017, \apj, 834, 52

\bibitem[{{Greene} {et~al.}(2016){Greene}, {Seth}, {Kim}, {L{\"a}sker}, {Goulding}, {Gao}, {Braatz}, {Henkel}, {Condon}, {Lo}, \& {Zhao}}]{Greene2016}
{Greene}, J.~E., {Seth}, A., {Kim}, M., {et~al.} 2016, \apjl, 826, L32

\bibitem[{{Greenhill} {et~al.}(2003){Greenhill}, {Booth}, {Ellingsen}, {Herrnstein}, {Jauncey}, {McCulloch}, {Moran}, {Norris}, {Reynolds}, \& {Tzioumis}}]{Greenhill2003}
{Greenhill}, L.~J., {Booth}, R.~S., {Ellingsen}, S.~P., {et~al.} 2003, \apj, 590, 162

\bibitem[{{Heckman} \& {Best}(2014)}]{Hechman2014}
{Heckman}, T.~M. \& {Best}, P.~N. 2014, \araa, 52, 589

\bibitem[{{Henkel} {et~al.}(2002){Henkel}, {Braatz}, {Greenhill}, \& {Wilson}}]{Henkel2002}
{Henkel}, C., {Braatz}, J.~A., {Greenhill}, L.~J., \& {Wilson}, A.~S. 2002, \aap, 394, L23

\bibitem[{{Herrnstein} {et~al.}(1997){Herrnstein}, {Moran}, {Greenhill}, {Inoue}, {Nakai}, {Miyoshi}, \& {Diamond}}]{Herrnstein1997}
{Herrnstein}, J., {Moran}, J., {Greenhill}, L., {et~al.} 1997, in American Astronomical Society Meeting Abstracts, Vol. 191, American Astronomical Society Meeting Abstracts, 25.07

\bibitem[{{H{\"o}gbom}(1974)}]{hoegbom74}
{H{\"o}gbom}, J.~A. 1974, \aaps, 15, 417

\bibitem[{{Kamali} {et~al.}(2017){Kamali}, {Henkel}, {Brunthaler}, {Impellizzeri}, {Menten}, {Braatz}, {Greene}, {Reid}, {Condon}, {Lo}, {Kuo}, {Litzinger}, \& {Kadler}}]{Kamali2017}
{Kamali}, F., {Henkel}, C., {Brunthaler}, A., {et~al.} 2017, \aap, 605, A84

\bibitem[{{Kamali} {et~al.}(2019){Kamali}, {Henkel}, {Koyama}, {Kuo}, {Condon}, {Brunthaler}, {Reid}, {Greene}, {Menten}, {Impellizzeri}, {Braatz}, {Litzinger}, \& {Kadler}}]{Kamali2019}
{Kamali}, F., {Henkel}, C., {Koyama}, S., {et~al.} 2019, \aap, 624, A42

\bibitem[{Karttunen {et~al.}(2017)Karttunen, Kr{\"o}ger, Oja, Poutanen, \& Donner}]{Karttunen2017}
Karttunen, H., Kr{\"o}ger, P., Oja, H., Poutanen, M., \& Donner, K.~J. 2017, Photometric Concepts and Magnitudes, ed. H.~Karttunen, P.~Kr{\"o}ger, H.~Oja, M.~Poutanen, \& K.~J. Donner (Berlin, Heidelberg: Springer Berlin Heidelberg), 91--102

\bibitem[{{Keimpema} {et~al.}(2015){Keimpema}, {Kettenis}, {Pogrebenko}, {Campbell}, {Cim{\'o}}, {Duev}, {Eldering}, {Kruithof}, {van Langevelde}, {Marchal}, {Molera Calv{\'e}s}, {Ozdemir}, {Paragi}, {Pidopryhora}, {Szomoru}, \& {Yang}}]{keimpema2015}
{Keimpema}, A., {Kettenis}, M.~M., {Pogrebenko}, S.~V., {et~al.} 2015, Experimental Astronomy, 39, 259

\bibitem[{Kuo {et~al.}(2010)Kuo, Braatz, Condon, Impellizzeri, Lo, Zaw, Schenker, Henkel, Reid, \& Greene}]{Kuo_2010}
Kuo, C.~Y., Braatz, J.~A., Condon, J.~J., {et~al.} 2010, The Astrophysical Journal, 727, 20

\bibitem[{{Kuo} {et~al.}(2011){Kuo}, {Braatz}, {Condon}, {Impellizzeri}, {Lo}, {Zaw}, {Schenker}, {Henkel}, {Reid}, \& {Greene}}]{kuo2011}
{Kuo}, C.~Y., {Braatz}, J.~A., {Condon}, J.~J., {et~al.} 2011, \apj, 727, 20

\bibitem[{{Kuo} {et~al.}(2020){Kuo}, {Braatz}, {Impellizzeri}, {Gao}, {Pesce}, {Reid}, {Condon}, {Kamali}, {Henkel}, \& {Greene}}]{Kuo2020}
{Kuo}, C.~Y., {Braatz}, J.~A., {Impellizzeri}, C.~M.~V., {et~al.} 2020, \mnras, 498, 1609

\bibitem[{{Kuo} {et~al.}(2013){Kuo}, {Braatz}, {Reid}, {Lo}, {Condon}, {Impellizzeri}, \& {Henkel}}]{Kuo2013}
{Kuo}, C.~Y., {Braatz}, J.~A., {Reid}, M.~J., {et~al.} 2013, \apj, 767, 155

\bibitem[{{Masini} {et~al.}(2022){Masini}, {Celotti}, \& {Campitiello}}]{Masini2022}
{Masini}, A., {Celotti}, A., \& {Campitiello}, S. 2022, \aap, 658, A68

\bibitem[{{Merloni} {et~al.}(2003){Merloni}, {Heinz}, \& {di Matteo}}]{merloni03}
{Merloni}, A., {Heinz}, S., \& {di Matteo}, T. 2003, \mnras, 345, 1057

\bibitem[{{Morishima} {et~al.}(2023){Morishima}, {Sudou}, {Yamauchi}, {Taniguchi}, \& {Nakai}}]{Morishima2023}
{Morishima}, Y., {Sudou}, H., {Yamauchi}, A., {Taniguchi}, Y., \& {Nakai}, N. 2023, \pasj, 75, 71

\bibitem[{National Radio Astronomy~Observatory(2020)}]{calibratori}
National Radio Astronomy~Observatory, N. 2020, The VLBA Calibrator List

\bibitem[{{Panessa} {et~al.}(2006){Panessa}, {Bassani}, {Cappi}, {Dadina}, {Barcons}, {Carrera}, {Ho}, \& {Iwasawa}}]{Panssa2006}
{Panessa}, F., {Bassani}, L., {Cappi}, M., {et~al.} 2006, \aap, 455, 173

\bibitem[{Pariev \& Bromley(1998)}]{Pariev1998}
Pariev, V.~I. \& Bromley, B.~C. 1998, The Astrophysical Journal, 508, 590

\bibitem[{{Peck} {et~al.}(2003){Peck}, {Henkel}, {Ulvestad}, {Brunthaler}, {Falcke}, {Elitzur}, {Menten}, \& {Gallimore}}]{Peck2003}
{Peck}, A.~B., {Henkel}, C., {Ulvestad}, J.~S., {et~al.} 2003, \apj, 590, 149

\bibitem[{{Pesce} {et~al.}(2015{\natexlab{a}}){Pesce}, {Braatz}, {Condon}, {Gao}, {Henkel}, {Litzinger}, {Lo}, \& {Reid}}]{Pescegbt2015}
{Pesce}, D.~W., {Braatz}, J.~A., {Condon}, J.~J., {et~al.} 2015{\natexlab{a}}, \apj, 810, 65

\bibitem[{{Pesce} {et~al.}(2015{\natexlab{b}}){Pesce}, {Braatz}, {Condon}, {Gao}, {Henkel}, {Litzinger}, {Lo}, \& {Reid}}]{Pesce2015}
{Pesce}, D.~W., {Braatz}, J.~A., {Condon}, J.~J., {et~al.} 2015{\natexlab{b}}, \apj, 810, 65

\bibitem[{{Pesce} {et~al.}(2023){Pesce}, {Braatz}, {Henkel}, {Humphreys}, {Impellizzeri}, \& {Kuo}}]{Pesce183ghz2023}
{Pesce}, D.~W., {Braatz}, J.~A., {Henkel}, C., {et~al.} 2023, \apj, 948, 134

\bibitem[{{Pesce} {et~al.}(2020{\natexlab{a}}){Pesce}, {Braatz}, {Reid}, {Condon}, {Gao}, {Henkel}, {Kuo}, {Lo}, \& {Zhao}}]{Pesce2020_geomdist}
{Pesce}, D.~W., {Braatz}, J.~A., {Reid}, M.~J., {et~al.} 2020{\natexlab{a}}, \apj, 890, 118

\bibitem[{{Pesce} {et~al.}(2020{\natexlab{b}}){Pesce}, {Braatz}, {Reid}, {Riess}, {Scolnic}, {Condon}, {Gao}, {Henkel}, {Impellizzeri}, {Kuo}, \& {Lo}}]{Pesce2020_H0}
{Pesce}, D.~W., {Braatz}, J.~A., {Reid}, M.~J., {et~al.} 2020{\natexlab{b}}, \apjl, 891, L1

\bibitem[{{Reid} {et~al.}(2009){Reid}, {Braatz}, {Condon}, {Greenhill}, {Henkel}, \& {Lo}}]{Reid2009}
{Reid}, M.~J., {Braatz}, J.~A., {Condon}, J.~J., {et~al.} 2009, \apj, 695, 287

\bibitem[{{Reid} {et~al.}(2013){Reid}, {Braatz}, {Condon}, {Lo}, {Kuo}, {Impellizzeri}, \& {Henkel}}]{Reid2013}
{Reid}, M.~J., {Braatz}, J.~A., {Condon}, J.~J., {et~al.} 2013, \apj, 767, 154

\bibitem[{{Reid} \& {Honma}(2014)}]{Reid2014}
{Reid}, M.~J. \& {Honma}, M. 2014, \araa, 52, 339

\bibitem[{{Schulz} {et~al.}(1995){Schulz}, {Muecke}, {Boer}, {Dresen}, \& {Schmidt-Kaler}}]{Schulz1995}
{Schulz}, H., {Muecke}, A., {Boer}, B., {Dresen}, M., \& {Schmidt-Kaler}, T. 1995, \aaps, 109, 523

\bibitem[{{Tarchi}(2012)}]{Tarchi2012}
{Tarchi}, A. 2012, in Cosmic Masers - from OH to H0, ed. R.~S. {Booth}, W.~H.~T. {Vlemmings}, \& E.~M.~L. {Humphreys}, Vol. 287, 323--332

\bibitem[{{Tarchi} {et~al.}(2011){Tarchi}, {Castangia}, {Henkel}, {Surcis}, \& {Menten}}]{Tarchi2011_KiloMaser}
{Tarchi}, A., {Castangia}, P., {Henkel}, C., {Surcis}, G., \& {Menten}, K.~M. 2011, \aap, 525, A91

\bibitem[{{Terashima} {et~al.}(2002){Terashima}, {Iyomoto}, {Ho}, \& {Ptak}}]{Terashima2002}
{Terashima}, Y., {Iyomoto}, N., {Ho}, L.~C., \& {Ptak}, A.~F. 2002, \apjs, 139, 1

\bibitem[{{Zhao} {et~al.}(2018){Zhao}, {Braatz}, {Condon}, {Lo}, {Reid}, {Henkel}, {Pesce}, {Greene}, {Gao}, {Kuo}, \& {Impellizzeri}}]{Zhao2018}
{Zhao}, W., {Braatz}, J.~A., {Condon}, J.~J., {et~al.} 2018, \apj, 854, 124

\bibitem[{{Zhu} {et~al.}(2011){Zhu}, {Zaw}, {Blanton}, \& {Greenhill}}]{zhu2011}
{Zhu}, G., {Zaw}, I., {Blanton}, M.~R., \& {Greenhill}, L.~J. 2011, \apj, 742, 73

\end{thebibliography}
\begin{appendix}

\section{Tentative maser features}
\label{apdx}
\begin{figure}
\includegraphics[width = 8.5cm]{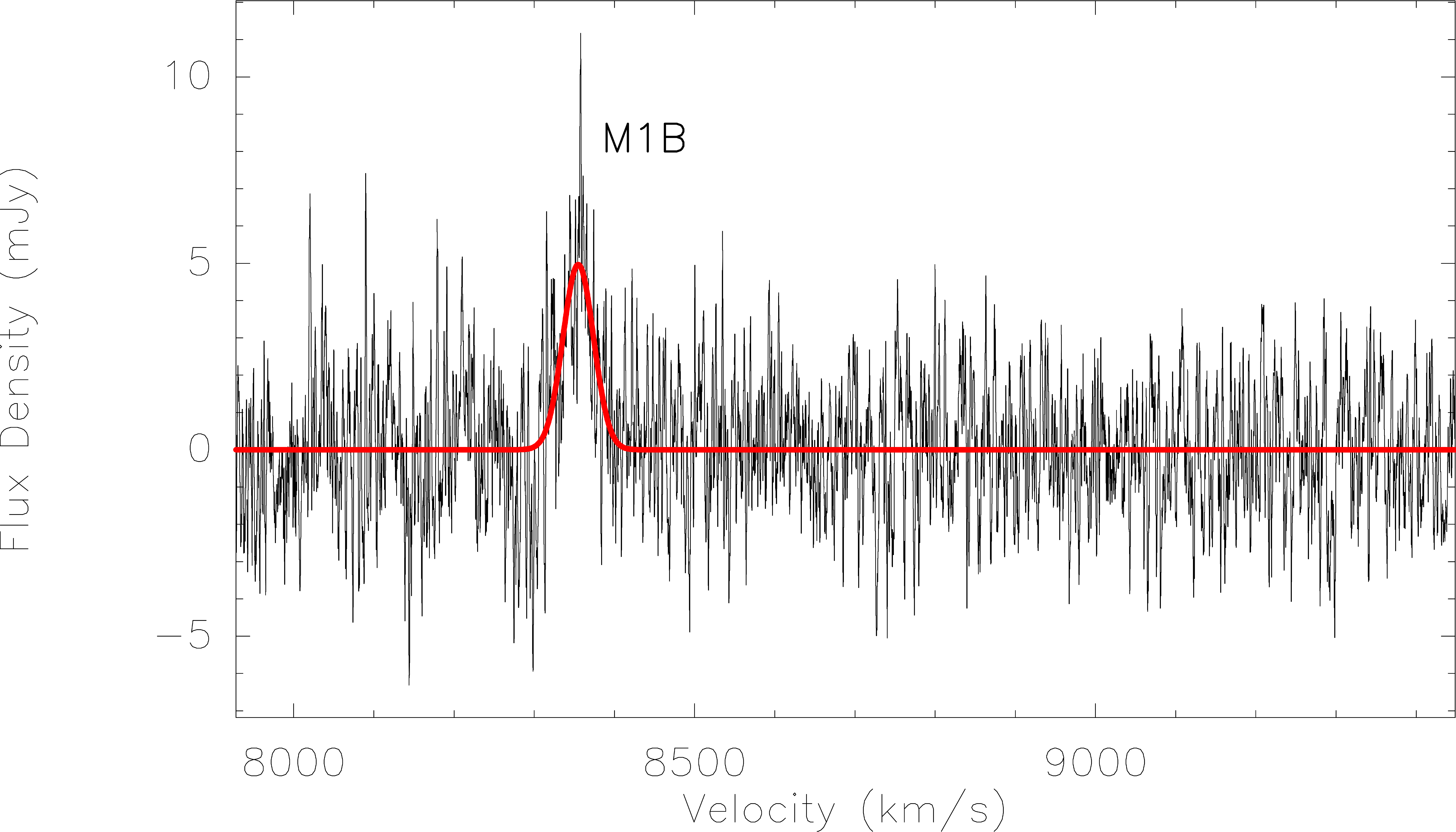}
\includegraphics[width = 8.5cm]{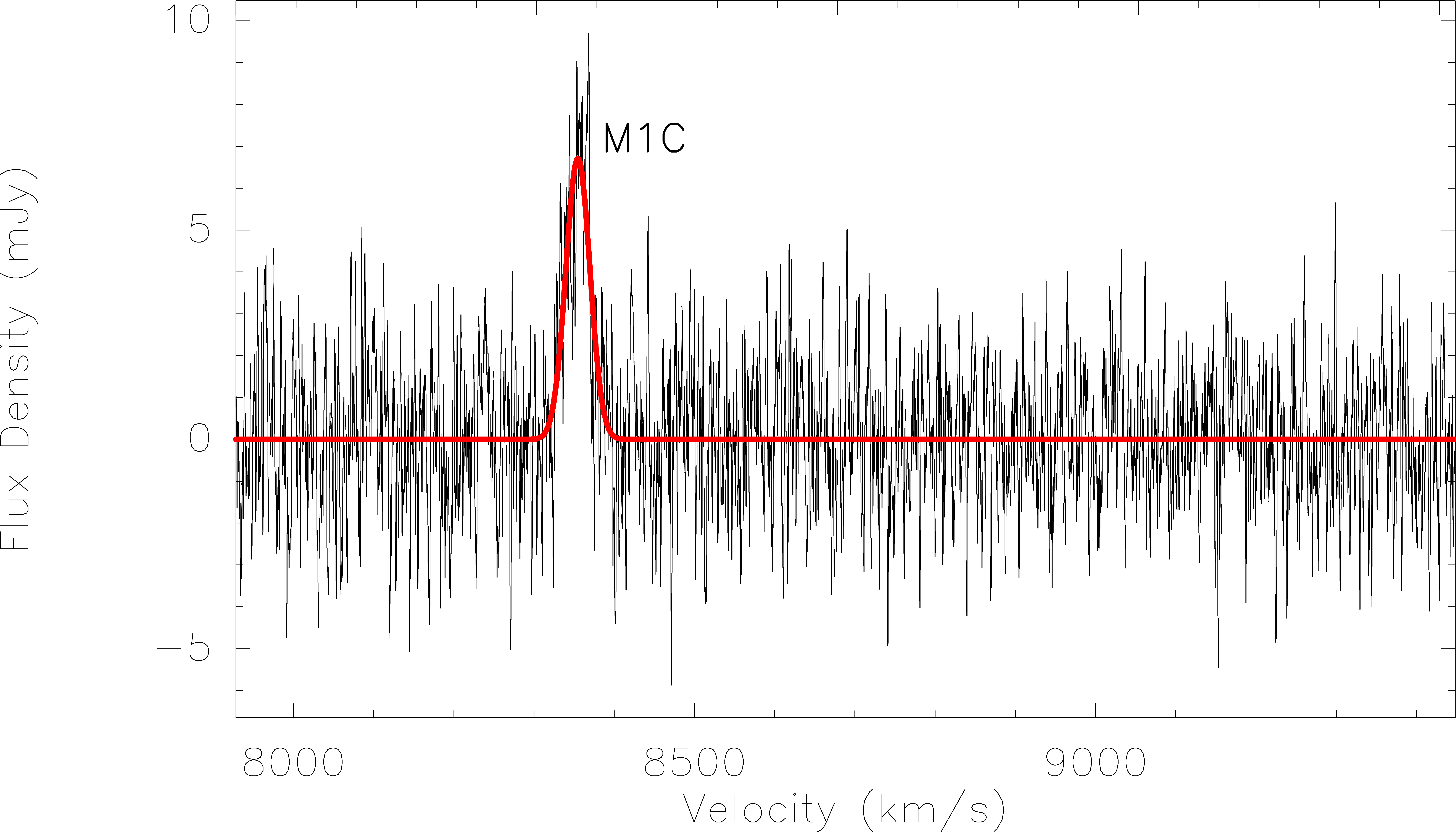}
\includegraphics[width = 8.5cm]{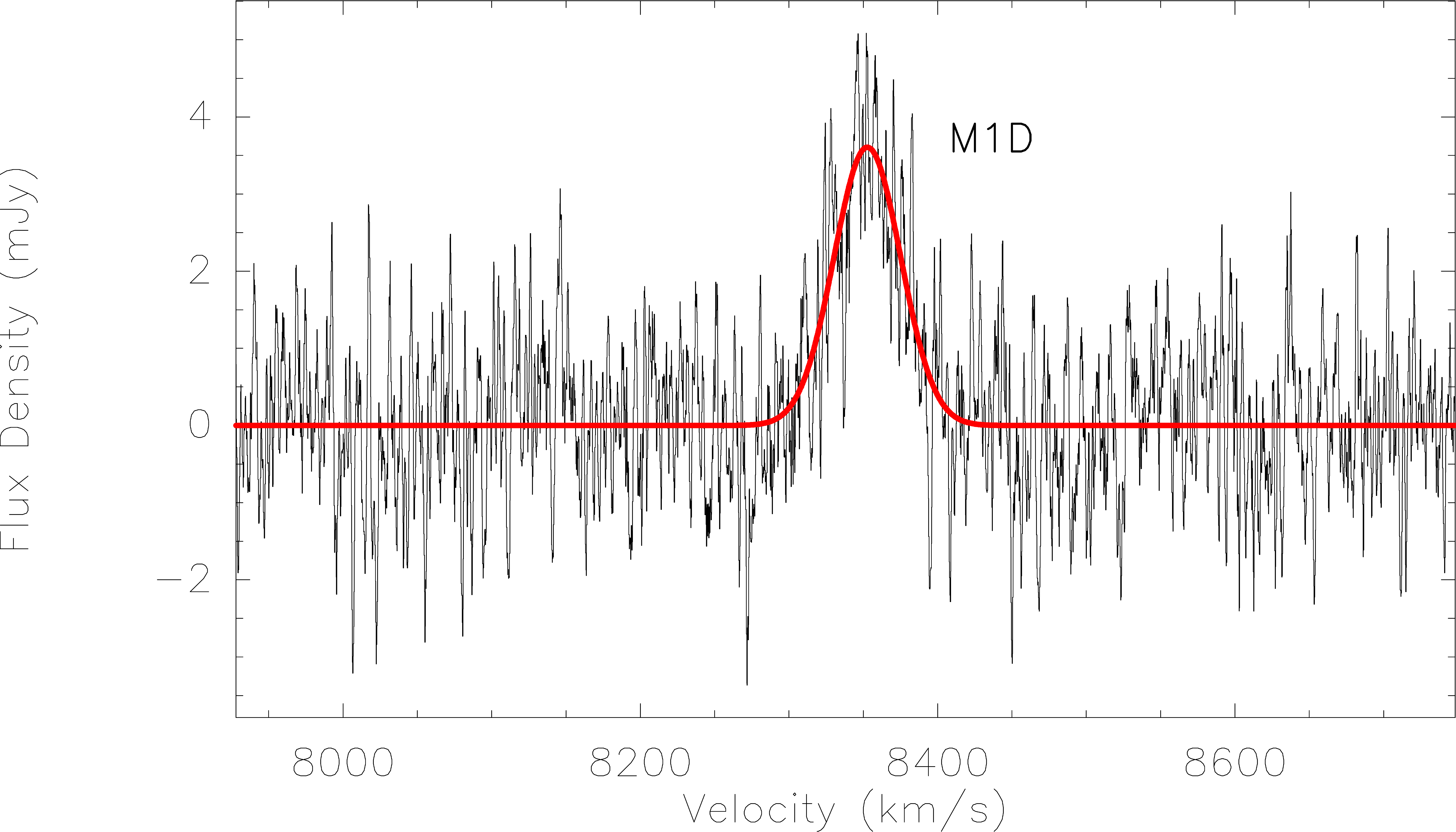}
\caption{ Spectrum with smooth \num{8}  with the corresponding Gaussian fit  (thin red line) and the  respective label of the tentative maser features observed in the epoch VLBA 2018.16 and VLBA 2018.83. 
}
\label{fig:M1Bfit}
\end{figure}

\begin{figure}
\includegraphics[width = 8.5cm]{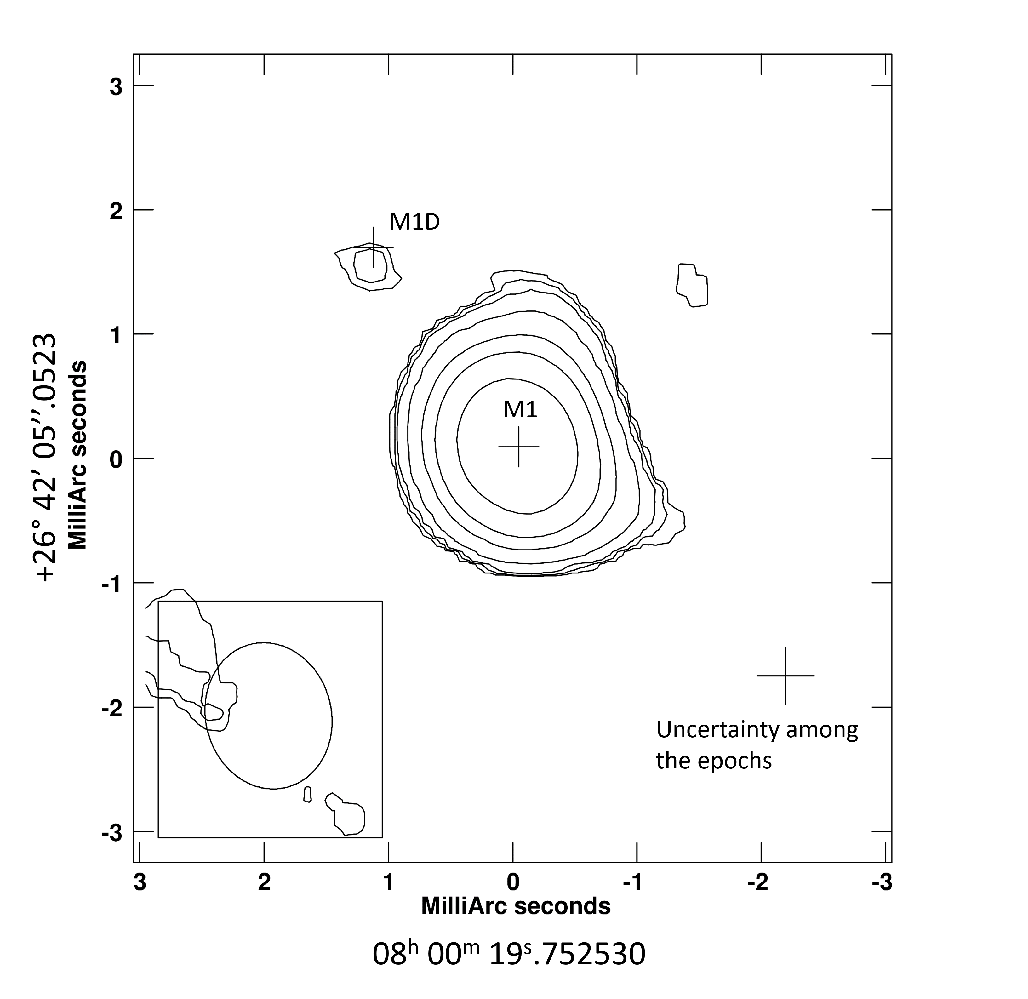}
\includegraphics[width = 8cm]{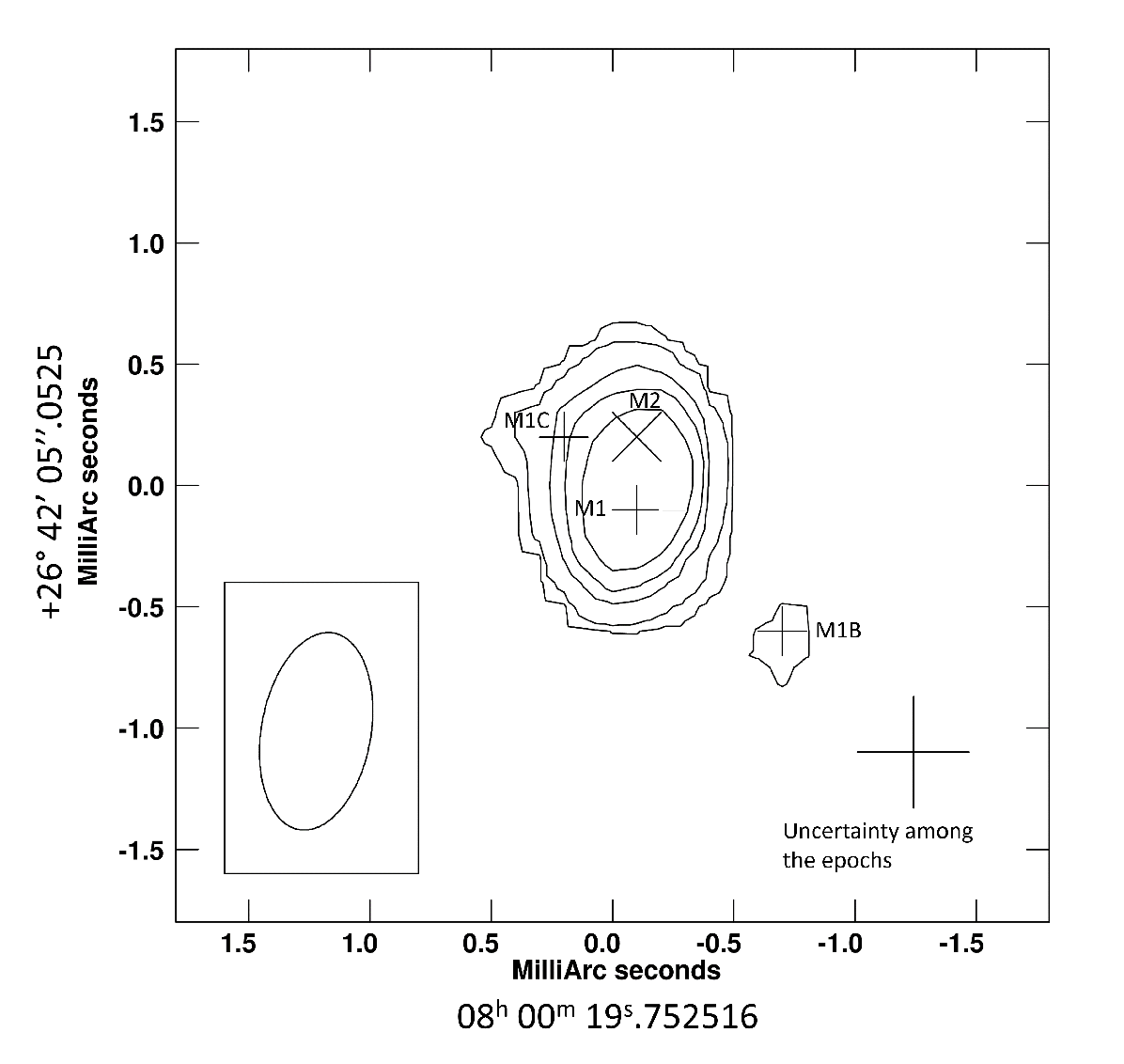}
\caption{ Contour maps of the zeroth moment with a cut at $2 \, \sigma$ of 
M1 observed in epoch 2018.11 (top) and in epoch 2018.83 (bottom). 
The positions of the tentative features (M1B-M1D) are indicated with plus symbols together with M1 (``$+$'' symbol) and M2 (``$\times$'' symbol).
}
\label{fig:Mom_tentative}
\end{figure}

In this appendix, we report some tentative features observed in the epochs VLBA 2018.83 and VLBA 2018.16. 
We define ``tentative'' a feature detected with a S$/$N $<$ \num{5}$\sigma$ that, however, is considered of interest for its proximity to the component M\num{1} (within $\sim \SI{2}{\mas} \simeq \SI{1.2}{\pc}$) and thus may provide useful clues to yield a complete picture of the water maser nature in IC\,485. All these maser features (named M1B, M1C, and M1D) have been detected with velocities close to the systemic one.
The most relevant seem to be the features M\num{1}B and M\num{1}D that have been identified both in the sqashed map and in the image cube (see Sect.~\ref{sec:obs_spect_vlba}). 
The spectra of each tentative maser feature, with a smooth function (\num{8} channel), are showed in Fig.~\ref{fig:M1Bfit} 
 where the Gaussian fit is also reported. 
The feature M\num{1}C (at \SI{8355(1)}{\km \per \s})  
seems strongly linked to the component M\num{1} \SI{8354.8(5)}{\km \per \s} due to the similar velocities of both. 
  The features M\num{1}B, M\num{1}C, were detected in the epoch VLBA 2018.83. M\num{1}D was detected in the epoch VLBA 2018.16 a velocity of \SI{8352.8(6)}{\km \per \s}.
In table~\ref{tab:masertentative} are reported the tentative features' parameters and in the Fig.~\ref{fig:Mom_tentative} are reported their position w.r.t the main maser emission M1 in the zeroth moment map of this latter.
The presence of these tentative detections displaced from the disk structures may  suggest a jet or an outflow maser and a possible composite maser's nature, already known and observed in the literature (e.g., NGC\num{1068},~\citealt{Gallimore1996}).

\begin{table*}
\caption{Parameters of the tentative maser features detected in IC\num{485}.}  
\label{tab:masertentative}
\centering
\begin{tabular}{lcccccc}
\hline \hline
  Maser  &    RA       &   Dec        &  Peak flux density     &       FWHM         &      Velocity-integrated                     & Peak velocity   \\
         &              &              &                        &                   &       flux density                         &                   \\  
          & ($\rm{^{h}:~^{m}:~^{s}}$) & ($\rm{^{\circ}:\,':\,''}$)   & ( \si{\mJy} )   &  (\si{\km\per\s})  & (\si{\mJy \cdot \km \, \s^{-1}}) & (\si{\km\per\s}) \\
\hline
M\num{1}B & 08:00:19.752470 & $+$26:42:05.0520 & \num{5.0(8)} & \num{45(6)} & \num{236(23)} & \num{8355(2)}  \\
M\num{1}C & 08:00:19.752537 & $+$26:42:05.0528 & \num{6.7(8)} & \num{35(3)} & \num{250(20)} & \num{8355(1)}  \\
M\num{1}D & 08:00:19.752617 & $+$26:42:05.0540 & \num{3.4(1)} & \num{48(1)} & \num{174.1(4)} & \num{8352.8(6)} \\ 
\hline
\end{tabular}
\tablefoot{The columns indicate the maser feature name, right ascension, declination obtain with the verb  ``imstat'' of AIPS; peak flux, the full width at half maximum, the area, i.e the velocity-integrated flux density,  and the peak velocity obtain by fitting them in CLASS.}
\end{table*}

\end{appendix}

\end{document}